\documentclass[iop]{emulateapj}

\newcommand{\kms}{\rm km~s^{-1}}
\newcommand{\kmsmpc}{\rm km~s^{-1}~Mpc^{-1}}
\newcommand{\dn}{D_{n}4000}

\slugcomment{\bf Last updated:\today}

\usepackage{color}
\usepackage{multirow}

\begin{document}

\title{A Complete Spectroscopic Census of Abell 2029: A Tale of Three Histories}

\author{Jubee Sohn$^{1}$, 
        Margaret J. Geller$^{1}$, H. Jabran Zahid$^{1}$, Daniel G. Fabricant$^{1}$}

\affil{$^{1}$ Smithsonian Astrophysical Observatory, 60 Garden Street, Cambridge, MA 02138, USA}

\begin{abstract}
A rich spectroscopic census of members of the local massive cluster Abell 2029 
includes 1215 members of A2029 and its two infalling groups, A2033 and a Southern Infalling Group (SIG).
The two infalling groups are identified in spectroscopic, X-ray and weak lensing maps. 
We identify active galactic nuclei (AGN), star-forming galaxies, E+A galaxies, and quiescent galaxies 
 based the spectroscopy.
The fractions of AGN and post-starburst E+A galaxies in A2029 are similar to those of other clusters. 
We derive the stellar mass ($M_{*}$)-metallicity of A2029 based on 227 star-forming members; 
 A2029 members within $10^{9} M_{\odot} < M_{*} < 10^{9.5} M_{\odot}$ are more metal rich than SDSS galaxies within the same mass range. 
We utilize the spectroscopic index $\dn$, a strong age indicator, to trace past and future evolution of the A2029 system. 
The median $\dn$ of the members decreases as 
 the projected clustercentric distance increases for all three subsystems. 
The $\dn - M_{*}$ relations of the members in A2029 and its two infalling groups differ significantly
 indicating the importance of stochastic effects for understanding the evolution of cluster galaxy populations.
In the main cluster, 
 an excess around $\dn \sim 1.8$ indicates that some A2029 members became quiescent galaxies 2-3 Gyr ago 
 consistent with the merger epoch of the X-ray sloshing pattern. 
\end{abstract}
           
\section{INTRODUCTION}

Galaxy clusters probe the effects of high density environments on galaxy evolution. 
\citet{Hubble31} recognized that cluster regions contain earlier types of galaxies. 
Many recent studies investigate differences among galaxy populations in clusters and lower density regions
 \citep{Dressler80, Balogh99, Rines05, Boselli06, Haines13, Haines15, Paccagnella16}.
In general, cluster galaxies are red, old, and quiescent 
 compared with their counterparts in the low density regions. 
This environmental dependence on the galaxy properties is expected in the hierarchical structure formation model; 
 cluster galaxies formed and evolved earlier in denser, more massive halos. 

Galaxy evolution in the cluster environment is complex;
 interactions with the intracluster medium (ICM), with other cluster member galaxies, 
 and with the global tidal field of the cluster are important. 
When a galaxy falls into the cluster environment, 
 interaction with the ICM gas may remove gas from individual galaxies
 (e.g. \citealp{Gunn72, Larson80, Moore96}). 
The impact of cluster environment depends on galaxy properties, including mass, morphology, and gas content 
 (e.g. \citealp{Christlein05, Blanton09, Peng10}). 
Thus, a large and complete sample of members is required 
 to reduce systematic effects on the study of cluster galaxy evolution.

Early studies identified cluster members based on photometric galaxy catalogs (e.g. \citealp{Oemler74}).
Alternative methods include control field subtraction \citep{Paolillo01} and red sequence selection \citep{DePropris98}. 
Red sequence selection is widely used 
 for identifying clusters from huge galaxy surveys \citep{Hao10, Rykoff14, Oguri14}. 
However, cluster member catalogs built with this approach are often contaminated by foreground/background interlopers \citep{Sohn18a}.

Dense spectroscopic surveys are critical for cluster member selection and for spectroscopic typing. 
Many previous studies use large stacked samples of spectroscopically identified cluster members 
 derived from several superimposed clusters (e.g. \citealp{Lewis02b, Haines13, Haines15, Paccagnella16}).  
The use of stacked cluster samples may introduce systematic issues 
 originating from the diversity of cluster properties 
 (e.g. redshift, mass, dynamical status of the sample clusters). 
 
A dense and complete spectroscopic survey of a single galaxy cluster 
 resolves some of these systematic issues. 
Recent studies examine cluster galaxy properties 
 based on a few hundred spectroscopic members identified in dense spectroscopic surveys
 \citep{Rines02, Smith12, Tyler13, Hwang12b, Geller14, Lee15, Sohn17a, Deshev17, Habas18}.
These spectroscopic surveys enable studies of cluster galaxy populations 
 (e.g. quiescent/star-forming, active galactic nuclei, and post-starburst galaxies)
 as well as the statistical distribution of cluster members 
 (e.g. luminosity, stellar mass, and the central velocity dispersion functions, \citealp{Rines08, Agulli14, Agulli16, Sohn17a, Song17}). 
More importantly, 
 this approach allows the properties of the galaxy clusters to be connected to their member properties
 \citep{Hwang09, Deshev17}. 

Abell 2029 is a massive cluster at $z = 0.078$ with an unusually rich spectroscopic survey
 \citep{Tyler13, Sohn17a, Sohn18b}. 
This cluster is one of the best sampled clusters in the universe. 
Moreover, A2029 has been studied based on multi-wavelength photometry, spectroscopy, weak lensing, and X-ray observations
 \citep{Clarke04, Bai07, Hicks10, Walker12, Tyler13, PaternoMahler13}. 
Based on the extensive dataset, 
 \citet{Tyler13} examine the star formation activity of the member galaxies. 
\citet{Sohn17a} measure the luminosity, stellar mass, and central velocity dispersion functions of the member galaxies. 

The dynamical evolution of A2029 is complex.  
A2029 has a distinctive X-ray sloshing pattern identified from high resolution {\it Chandra} X-ray imaging
 \citep{Clarke04, PaternoMahler13}. 
Comparison between the X-ray feature and hydrodynamic simulations suggests that 
 A2029 experienced the accretion of a subcluster 2-3 Gyr ago \citep{PaternoMahler13}.
Furthermore, A2029 is currently massively accreting. 
\citet{Sohn18b} investigate the structure of the A2029 system 
 based on multi-wavelength probes including the number density map of spectroscopic members, weak lensing and X-ray maps. 
They identify at least two subsystems, A2033 and SIG, within the infall region of A2029. 
These two subsystems will probably be accreted onto A2029 within a few Gyr. 
This complicated dynamical history and future make A2029  
 an interesting target for studying the connection between accretion and galaxy populations.  

Here we examine the census of spectroscopic properties of A2029 member galaxies.  
We use a complete sample of $\sim 1200$ spectroscopic members in the A2029 system 
 including $\sim 50-70$ members in each of the two infalling groups. 
This rich dataset enables a comparative study of galaxy populations in the cluster and its infalling groups. 
Combining the spectroscopic properties of the galaxy populations with the X-ray structure of A2029, 
 we connect galaxy evolution in A2029 to its accretion history. 
Furthermore, currently differing galaxy populations in A2029, A2033, and SIG 
 suggest diverse consequences of subcluster accretion on the resultant cluster population. 

We describe the data in Section \ref{data}. 
We describe the spectroscopic member selection and 
 physical properties of A2029 members in Section \ref{spec}. 
In Section \ref{census}, 
 we examine the spectroscopic census of the A2029 population; we identify
 Active Galactic Nuclei (AGN), star-forming, post-starbust E+A, and quiescent galaxies.  
We connect evolution of the galaxy population in the cluster to past and future accretion of substructures in Section \ref{discussion}. 
We conclude in Section \ref{conclusion}. 
We assume a standard cosmology of $H_{0} = 70~\kmsmpc$, $\Omega_{m} = 0.3$, and $\Omega_{\Lambda} = 0.7$ throughout. 

\section{Data}\label{data}

Abell 2029 is one of the best sampled clusters in the local universe \citep{Tyler13, Sohn17a, Sohn18b}. 
The intensive redshift survey from \citet{Sohn18b} includes 1215 spectroscopic members 
 within $R_{cl} < 40\arcmin$ of the cluster center. 
Only Coma has a comparable number of spectroscopically identified members. 
The large number of spectroscopic members
 enables studies of the detailed physical properties of members within a single massive cluster. 

We use the dense and complete A2029 spectroscopic survey from \citet{Sohn18b}, 
 which extends the surveys of \citet{Tyler13} and \citet{Sohn17a}. 
Here we briefly describe the spectroscopic survey. 
More details are in \citet{Sohn18b}. 
 
The basic photometric galaxy catalog is the Sloan Digital Sky Survey (SDSS) Data Release 12 (DR12).  
Extended sources with $probPSF = 0$ and brighter than $r = 22$ mag within $R_{cl} < 100\arcmin$ 
 are the targets for the redshift survey. 
Throughout this study, we use composite model (cModel) magnitudes after foreground extinction correction. 

We first compile redshifts from previous surveys including SDSS DR12. 
There are 3109 objects with SDSS redshifts.
We add 439 redshifts from NASA/IPAC Extragalactic Database (NED) 
 and one redshift from the 1.5m telescope on Mt. Hopkins \citep{Sohn15}.

Most of the spectra of A2029 galaxies were obtained with Hectospec mounted on the MMT 6.5 telescope. 
Hectospec is a fiber-fed multi-object spectrograph with a $\sim 1$ deg$^{2}$ field of view \citep{Fabricant05}. 
Using Hectospec, 
 \citet{Tyler13} measure the redshifts and H$\alpha$ equivalent widths of 1369 galaxies in the A2029 field. 
Based on these measurements, they identify cluster members and examine the evolution of star-forming galaxies in the cluster environment. 
We compile the spectra from \citet{Tyler13} through the MMT archive
 \footnote{http://oirsa.cfa.harvard.edu/archive/search/}. 
 
We carried out a deeper spectroscopic survey using MMT/Hectospec \citep{Sohn17a, Sohn18b}. 
These spectra of A2029 galaxies were also obtained using the 270 line mm$^{-1}$ Hectospec grating. 
The spectra of all targets uniformly cover 3800 - 9100 \AA~ with 6.2 \AA~ spectral resolution. 
The typical exposure time was an hour per field. 

To reduce the Hectospec data, 
 we use the IDL HSRED v2.0 package 
 \footnote{Originally developed by R.Cool and modified by the MMT Telescope Data Center.}. 
We use RVSAO \citep{Kurtz98} to measure the redshifts 
 based on cross-correlation of observed spectra with template spectra collected for this purpose \citep{Fabricant05}. 
We visually inspect the cross-correlation redshift measurements and 
 classify them into three groups: 
 `Q' for high-quality fits, `?' for ambiguous cases, and `X' for poor fits.  
We also classify the Hectospec spectra from \citet{Tyler13} in the same way. 
We obtain a total of 2704 high-quality redshifts from all of the Hectospec observations. 
The typical redshift uncertainty of Hectospec redshifts is $32~\kms$. 

The spectroscopic survey of A2029 is 90\% uniformly complete to $r = 20.5$ within $R_{cl} < 30\arcmin$ 
 (see Figure 1 in \citealp{Sohn18b}). 
The survey completeness decreases outside $R_{cl} = 30\arcmin$
 and is 67\% complete within $R_{cl} < 40\arcmin$. 

\section{Spectroscopic Properties of A2029 Cluster Member Galaxies}\label{spec}

Based on the extensive spectroscopic sample, 
 we explore the properties of galaxies in the A2029 region. 
We first identify cluster members, and foreground/background galaxies
 using the caustic technique (Section \ref{memsel}). 
For the spectroscopically identified cluster members, 
 we measure galaxy properties including 
 stellar mass (Section \ref{msest}), 
 $\dn$ (Section \ref{dnest}), 
 velocity dispersion ($\sigma$, Section \ref{sigma}), and 
 emission-line fluxes (Section \ref{flest}).

\subsection{Member Selection}\label{memsel}

We identify members of A2029 based on the caustic technique \citep{Diaferio97, Diaferio99, Serra13}. 
The caustic technique is a tool that derives the mass profile of a cluster 
 based on calculation of the escape velocity as a function of clustercentric radius. 
In deriving the mass profile, 
 the technique calculates the `trumpet-like' boundary of the cluster \citep{Kaiser87}; 
 the galaxies within this boundary (caustic pattern) are cluster members. 
The technique identifies $\sim95\%$ of cluster members within $3R_{200}$ 
 from mock catalogs that contain $\sim1000$ galaxies including $\sim 180$ members per cluster \citet{Serra13}. 
Only $\sim 8\%$ of identified members are contaminating interlopers within $3R_{200}$. 

In the relative rest-frame velocity difference versus projected distance domain, the R-v diagram, 
 the caustic pattern clearly separates cluster members from foreground/background galaxies
 (see Figure 2 in \citealp{Sohn18b}). 
We identify 1215 spectroscopic members of the A2029 systems within the caustics. 

\citet{Sohn18b} examine the structure of A2029 based on spectroscopy, weak lensing and X-ray maps. 
There are two subsystems identified in all three maps in the infall region of A2029: 
 Abell 2033 and a Southern Infalling Group (SIG). 
The members of these two subsystems are well within the A2029 caustics (Figure 4 in \citealp{Sohn18b}), 
 indicating that two subsystems are dynamically connected to A2029. 
Based on a two-body model \citep{Beers82}, 
 \citet{Sohn18b} suggest that these two subsystems are gravitationally bound to A2029 
 and will accrete onto A2029 within a few Gyr. 

Following \citet{Sohn18b}, 
 we refer to the A2029 system as the three components: A2029, A2033 and SIG. 
There are 1215 spectroscopic members in this A2029 system within $R_{proj} < 8.8$ Mpc, 
 where $R_{proj}$ is the projected distance from the center of A2029. 
We also identify members of A2033 and SIG within 
 $R_{proj, group} < 500$ kpc and $|\Delta cz| / (1 + z_{group}) < 2000~\kms$, 
 where $R_{proj, group}$ is the projected distance from the center of the two subsystems and 
 $z_{group}$ is the central redshift of A2033 or SIG. 
A2033 and SIG consist of 57 and 70 members, respectively. 
At the projected distance to A2033 and SIG, 
 the possible contamination by A2029 members is 8 and 18, respectively. 
We discuss the negligible impact of this contamination in Section \ref{dn}. 
Hereafter, we refer to the 57 and 70 members as A2033 and SIG members. 
The remaining 1088 members in the A2029 system belong to the central cluster, A2029, and its infall region.

\subsection{Stellar Mass}\label{msest}

We determine the stellar mass of each A2029 member
 using the Le PHARE fitting code \citep{Arnouts99, Ilbert06},
 which estimates a mass-to-light ratio
 based on $\chi^{2}$ synthetic spectral energy distribution fitting.
We compare the SDSS DR12 $ugriz$ cModel magnitudes of individual galaxies
 with the stellar population synthesis (SPS) models 
 generated by the \citet{Bruzual03} code, with a \citet{Chabrier03} IMF, and three metallicities.  
A set of synthetic SED models includes
 different star formation histories, foreground extinction, and stellar population ages. 
We assume an exponentially declining star forming history
 with an $e-$folding time scale $\tau =  0.1, 0.3, 1, 2, 3, 5, 10, 15, 30$. 
We also use the \citet{Calzetti00} extinction law with an E(B-V) range of 0.0 to 0.6 and 
 with stellar population ages between 0.01 to 13 Gyr. 
The Le PHARE code calculates the probability distribution function (PDF) for the stellar mass. 
Our estimated stellar mass is the median of the appropriate PDF. 

The typical absolute uncertainty in the stellar mass measured 
 from the SED fitting is $\sim 0.3$ dex \citep{Conroy09}. 
Uncertainties in star formation history, metallicity, dust extinction, 
 SED models, IMF, and the SED fitting method
 propagate to the uncertainty and systematic error in the stellar mass. 
The stellar mass estimate based on the Le PHARE code is systematically lower by $\sim 0.1$ dex than
 the mass estimates from other approaches \citep{Zahid14a}. 
Thus, our mass estimates are only relatively accurate within the typical statistical uncertainty. 

The stellar masses of the brightest galaxies of A2029 and A2033 we initially derive from SDSS DR12 are underestimated. 
Their SDSS DR12 cModel magnitudes are significantly overestimated: 
 $r= 17.29$ for the A2029 BCG and $r = 17.44$ for the A2033 BCG. 
Their DR7 cModel magnitudes are more reasonable: $r = 13.59$ for the A2029 BCG and $r = 14.20$ for the A2033 BCG, 
 similar to R-band magnitudes of these galaxies listed in NED. 
Therefore, we use the SDSS DR7 cModel magnitudes to estimate the stellar masses for these BCGs.  

\subsection{$\dn$}\label{dnest}

The $\dn$ index is the flux ratio between two spectral windows 
 around the 4000 \AA~ break \citep{Bruzual83, Balogh99}. 
We use the $\dn$ index as a marker of the stellar population age.  
Following the definition from \citet{Balogh99}, 
 we calculate the index 
 as a ratio between the flux in the interval 4000 - 4100 \AA~ and the flux in the interval 3850 - 3950 \AA. 
The $\dn$ indices for A2029 galaxies come directly from the spectra
 obtained in the SDSS, BOSS, and Hectospec observations. 
The $\dn$ values measured for the same objects from the Hectospec and SDSS spectra agree to within $\sim5\%$. 

We measure $\dn$ for essentially every member of the A2029 system (1198 members, $\sim 98.6\%$). 
Because of the completeness of the $\dn$ measurement, 
 investigations based on $\dn$ contain essentially no systematic bias. 
 
The $\dn$ index is often used to distinguish star-forming and quiescent galaxies 
 because of its bimodal distribution (e.g. \citealp{Kauffmann03, Woods10, Geller14}). 
This bimodality results from the fact that 
 the $\dn$ index is sensitive to the stellar population age. 
\citet{Sohn17a} identify quiescent galaxies in A2029 with $\dn > 1.5$
 to construct the velocity dispersion function. 
We follow this selection (e.g. see also \citealp{Woods10, Zahid16b}) 
 for identifying quiescent galaxies in A2029. 

\subsection{Velocity Dispersion}\label{sigma}

\citet{Sohn17a} publish central velocity dispersion measurements of A2029 quiescent galaxies with $\dn > 1.5$. 
They use central velocity dispersions from the Portsmouth reduction \citep{Thomas13} for the galaxies with SDSS spectroscopy. 
They also measure the central velocity dispersions for galaxies with Hectospec data. 
We follow their procedure to obtain central velocity dispersions of A2029 galaxies for this extended sample. 

We first compile velocity dispersions from the Portsmouth reduction \citep{Thomas13} 
 for the galaxies with SDSS spectroscopy. 
Because the velocity dispersions from the Portsmouth reduction are essentially identical to 
 those measured from Hectospec \citep{Fabricant13}, 
 we use these measurement without significant corrections. 
To measure the velocity dispersion from the SDSS spectra, 
 \citet{Thomas13} use the Penalized Pixel-Fitting (pPXF) code \citep{Cappellari04} and 
 the stellar population templates from \citet{Maraston11} that are based on the MILES stellar library \citep{SanchezBlazquez06}. 
The best-fit velocity dispersion is derived by comparing the spectra and the templates. 
From the Portsmouth reduction, 
 we obtain 416 velocity dispersions for A2029 members.
These velocity dispersions have a typical uncertainty of $7~\kms$. 

For galaxies with MMT/Hectospec spectra, 
 we measure velocity dispersions with the University of Lyon Spectroscopic analysis Software 
 (ULySS, \citealp{Koleva09}). 
ULySS derives the best-fit velocity dispersion based on a chi-square fit of the Hectospec spectra to the stellar population templates. 
We use stellar templates constructed with the PEGASE-HR code and the MILES stellar library. 
We convolved these stellar templates to the Hectospec resolution with various velocity dispersions. 
To minimize the uncertainty in the velocity dispersion measurements, 
 the fitting range is limited to the rest-frame spectral range 4100 - 5500 \AA~ \citep{Fabricant13}. 
We measure 765 velocity dispersions from Hectospec data. 
The median uncertainty of the Hectospec velocity dispersions is $\sim 17~\kms$. 

We apply an aperture correction to the velocity dispersion measurements 
 because SDSS and Hectospec obtain the spectra through $3\arcsec$ and $1.5\arcsec$ fibers, respectively.
We define the aperture correction following \citep{Zahid16}: 
\begin{equation}
\sigma_{\rm A} / \sigma_{\rm B} = (R_{\rm A} / R_{\rm B})^{\beta}. 
\end{equation}
We use 270 quiescent objects with both SDSS and Hectospec velocity dispersions
 within the range of $100 < \sigma < 450~\kms$ and $\Delta \sigma < 100~\kms$
 to determine $\beta$ for the aperture correction. 
The best fit parameter is $\beta = -0.059 \pm 0.014$, 
 consistent with $\beta = -0.046 \pm 0.013$ from \citet{Zahid16} and 
 $\beta = -0.054 \pm 0.005$ from \citet{Sohn17a}.

We quote the central velocity dispersion within a fiducial physical aperture of 3 kpc (rest-frame)
 following \citet{Zahid16} and \citet{Sohn17a}.
Throughout this paper, $\sigma$ indicates the central velocity dispersion within the 3 kpc aperture.
We use the velocity dispersion of quiescent galaxies with $\dn > 1.5$ where the random motion of stars dominates over stellar kinematics. 
Table \ref{memcat} lists the 140 velocity dispersions and $\dn$s of A2029 members 
 not included in \citet{Sohn17a}. 
Previous measurements for 834 A2029 members are included in Table 2 of \citet{Sohn17a}. 
 
\begin{deluxetable*}{lcccccccc}
\tablecolumns{9}
\tabletypesize{\footnotesize}
\setlength{\tabcolsep}{0.05in}
\tablecaption{Physical Properties of the Spectroscopic Members of A2029}
\tablehead{
\colhead{Object ID} & \colhead{R.A.}  & \colhead{Decl.} &
\colhead{z} & \colhead{$\log M_{*}$}  & \colhead{$\dn$} & \colhead{$\sigma$} & 
\colhead{spec. type\tablenotemark{a}} & \colhead{E+A?} \\
\colhead{}          & \colhead{(deg)} & \colhead{(deg)} &
\colhead{}  & \colhead{$M_{\odot}$}   & \colhead{}      & \colhead{($\kms$)} & 
\colhead{}                            & \colhead{}     }
\startdata
1237658780557836308 & 227.740259 &   5.766147 & $0.07731 \pm 0.00010$ & $ 10.90 \pm  0.04$ & $2.03 \pm 0.04$ & $281 \pm   5$ & N & N \\
1237658780557836338 & 227.737793 &   5.762320 & $0.07593 \pm 0.00007$ & $  8.22 \pm  0.33$ & $2.03 \pm 0.07$ & $105 \pm  14$ & N & N \\
1237658780557836305 & 227.744635 &   5.770809 & $0.07455 \pm 0.00007$ & $ 10.91 \pm  0.05$ & $2.14 \pm 0.04$ & $329 \pm   5$ & N & N \\
1237658780557836311 & 227.738249 &   5.754465 & $0.07726 \pm 0.00009$ & $ 10.20 \pm  0.06$ & $1.86 \pm 0.04$ & $238 \pm   9$ & N & N \\
1237658780557836336 & 227.732202 &   5.761856 & $0.08085 \pm 0.00014$ & $-99.00 \pm  0.00$ & $1.82 \pm 0.11$ & $ 84 \pm  31$ & N & N \\
1237658780557836342 & 227.749463 &   5.769346 & $0.07828 \pm 0.00007$ & $  8.65 \pm  0.19$ & $1.91 \pm 0.05$ & $ 49 \pm  17$ & N & N \\
1237658780557836316 & 227.735039 &   5.751555 & $0.07921 \pm 0.00008$ & $ 10.40 \pm  0.04$ & $2.08 \pm 0.04$ & $237 \pm   6$ & N & N \\
1237658780557836337 & 227.732491 &   5.765348 & $0.07899 \pm 0.00006$ & $  9.81 \pm  0.11$ & $1.88 \pm 0.05$ & $112 \pm  10$ & N & N \\
1237658780557836369 & 227.731678 &   5.764879 & $0.07735 \pm 0.00015$ & $-99.00 \pm  0.00$ & $1.56 \pm 0.09$ & $123 \pm  41$ & N & N
\enddata
\label{memcat}
\tablecomments{The entire table is available in machine-readable form in the online journal. 
Here, a portion is shown for guidance regarding its format. }
\end{deluxetable*} 
 
\subsection{Emission Line Flux}\label{flest}

We use emission line strengths 
 to study the physical properties including nuclear activity and metallicity 
 of the late-type galaxies in A2029. 
We first collect the line fluxes for the objects with SDSS spectra
 measured by the MPA/JHU Group \footnote{https://www.mpa.mpa$-$garching.mpg.de/SDSS/}
 and the Portsmouth Group 
 \footnote{https://www.sdss3.org/dr10/spectro/galaxy\_portsmouth.php} \citep{Thomas13}. 
The MPA/JHU and Portsmouth catalogs include flux measurements of 
 273 and 288 galaxies in the A2029 field, respectively. 
The line flux measurements from the two catalogs are consistent within $\sim 0.1$ dex
 ($\sim 0.2$ dex for [O II]$\lambda\lambda3726+3729$, \citealp{Thomas13}). 
Thus, the flux measurements from the MPA/JHU and Portsmouth catalogs are interchangeable. 
For 270 duplicated objects, 
 we take the MPA/JHU value for further analysis. 

We measure line fluxes for galaxies observed with Hectospec 
 same fitting procedure used in \citet{Zahid13}. 
The fitting procedure uses the MPFIT package \citep{Markwardt09} implemented in IDL. 
We fit the continuum of each galaxy with stellar population synthesis model of \citet{Bruzual03}. 
We fit each emission line in the continuum subtracted spectrum with a Gaussian and 
 use the best-fit parameters to derive the line flux. 
We calculate the observational uncertainties in the line flux measurements 
 by standard error propagation of the estimated uncertainties in the spectrum.

The flux measurements for the [O II]$\lambda\lambda3726+3729$ doublet may be uncertain 
 because the doublet is hardly resolved in the SDSS and Hectospec spectra \citep{Thomas13}. 
This [O II] line is essential for estimating the metallicity of emission-line galaxies
 (e.g. \citealp{Kobulnicky04}). 
Previous studies use the sum of the doublet 
 when deriving the metallicity based on the SDSS and/or Hectospec spectra 
 \citep{Zahid14a, Zahid14b, Wu17}. 
Following this procedure, 
 we also use the sum of the [O II] doublet 
 when we estimate the metallicity of A2029 galaxies (Section \ref{mzr}). 
We refer to the sum as [O II]$\lambda3727$ hereafter. 

We adopt a dust extinction correction based on the \citet{Cardelli89} extinction curve. 
We assume an intrinsic H$\alpha$/H$\beta$ ratio of 2.86 
 (case B recombination; \citealp{Osterbrock06}). 
The active galactic nuclei in A2029 may have a different intrinsic H$\alpha$/H$\beta$ ratio 
 (e.g. H$\alpha$/H$\beta \sim 3.0$, \citealp{Dong08, Baron16}). 
We apply AGN diagnostics (see Section \ref{agn}) with different intrinsic Balmer ratios
 to test the effect of the dust correction. 
The number of AGN does not change when we use a Balmer ratio in the range $2.86 - 3.10$. 
Thus, we simply use the Balmer ratio of 2.86 regardless of the type of object.  

\section{CENSUS OF A2029 MEMBER GALAXIES}\label{census}

Here we examine a census of A2029 cluster members
 based on the complete spectroscopic sample. 
The complete survey enables statistical analysis 
 of the galaxy populations in the cluster without significant sample selection bias.
The number of members is large enough to avoid the stacking techniques often used to explore cluster populations 
 \citep{Balogh99, Haines13, Paccagnella16}.
 
\begin{figure*}
\centering
\includegraphics[scale=1.0]{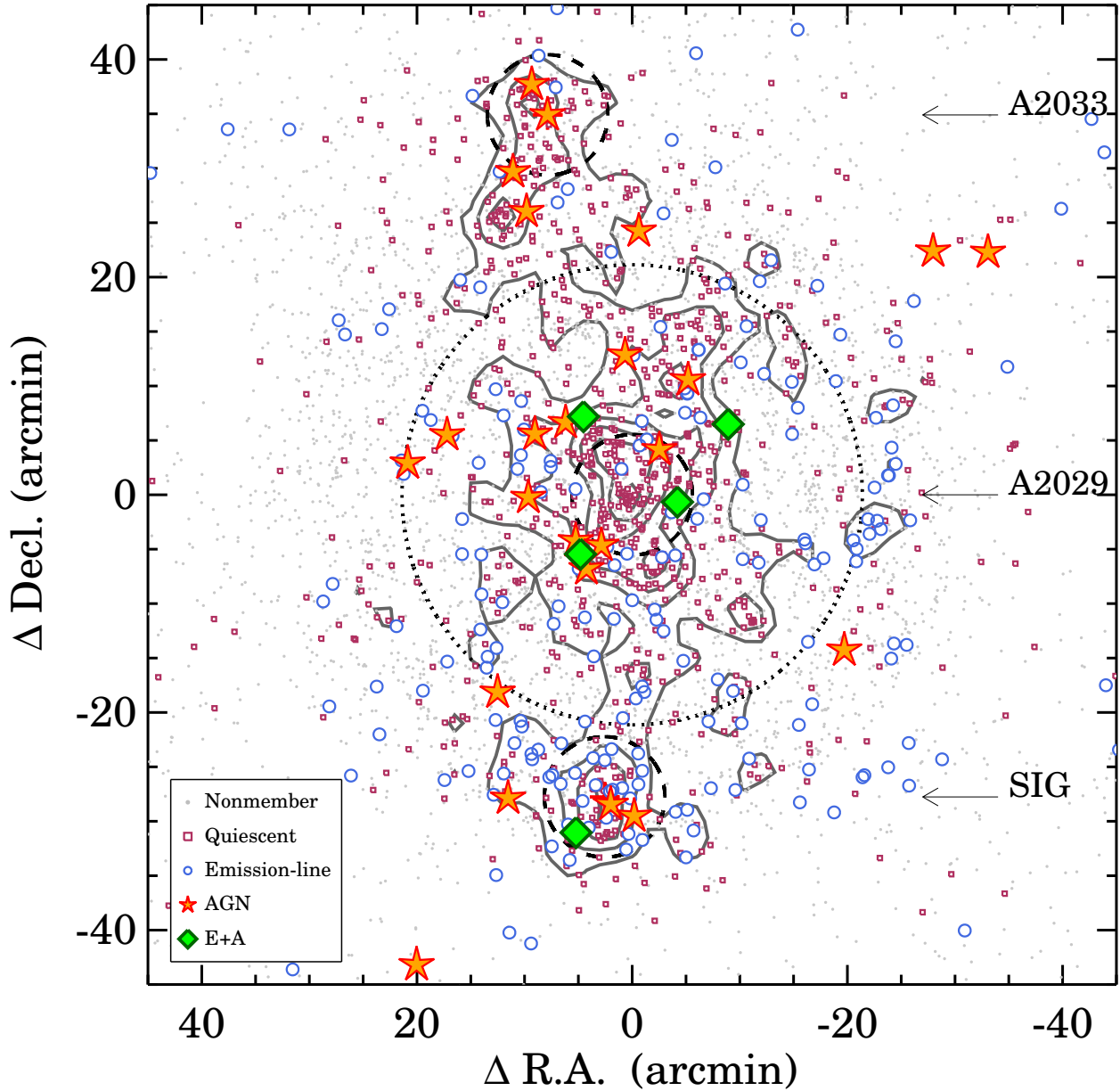}
\caption{Spatial distribution of galaxies in the A2029 field. 
Gray dots are spectroscopic targets. 
Red squares are quiescent ($\dn > 1.5$) A2029 members and blue circles are star-forming population ($\dn \leq 1.5$) in A2029. 
Red stars and green diamonds indicate AGNs and E+A galaxies, respectively. 
The underlying contour shows the number density of A2029 members. 
The labels and arrows mark the position (declination) of each system. 
The dotted circle centered on A2029 show $R_{cl} = R_{200} = 1.91$ Mpc. 
The dashed circles centered on A2033 (north), A2029 (center), and SIG (south) show $R_{cl} = 500$ kpc. }
\label{spatial}
\end{figure*} 

Figure \ref{spatial} summarizes the galaxy populations. 
Figure \ref{spatial} displays the spatial distribution of spectroscopic targets in the A2029 field. 
Gray dots are spectroscopically identified nonmembers of A2029. 
Blue circles and red squares indicate spectroscopic members 
 with $\dn > 1.5$ and $\dn \leq 1.5$, respectively. 
Gray contours show the number density of the spectroscopically selected A2029 members.  
A red cross marks the position of the brightest cluster galaxy (BCG, IC 1101). 
Red stars show active galactic nuclei (Section \ref{agn}).
We examine the stellar mass and metallicity relation 
 based on the members with emission-lines (blue circles, Section \ref{mzr}). 
Green diamonds mark 
 the position of post-starburst (E+A) galaxies (Section \ref{ea}). 
Finally, we discuss the $\dn$ distribution of the members of A2029 and of the infalling systems A2033 and SIG in Section \ref{dn}. 
 
\subsection{AGNs in A2029}\label{agn}

We adopt the widely used BPT diagnostic diagram \citep{Baldwin81} 
 to determine the spectral types of emission-line galaxies. 
We require line fluxes with $S/N > 3$
 for H$\alpha$, H$\beta$, [OIII]$\lambda5007$, and [NII]$\lambda6584$,
 for AGN diagnosis. 
There are 277 objects ($23 \pm 1\%$) with reliable emission-line fluxes 
 among 1215 A2029 system members. 
 
\begin{figure}
\centering
\includegraphics[scale=0.50]{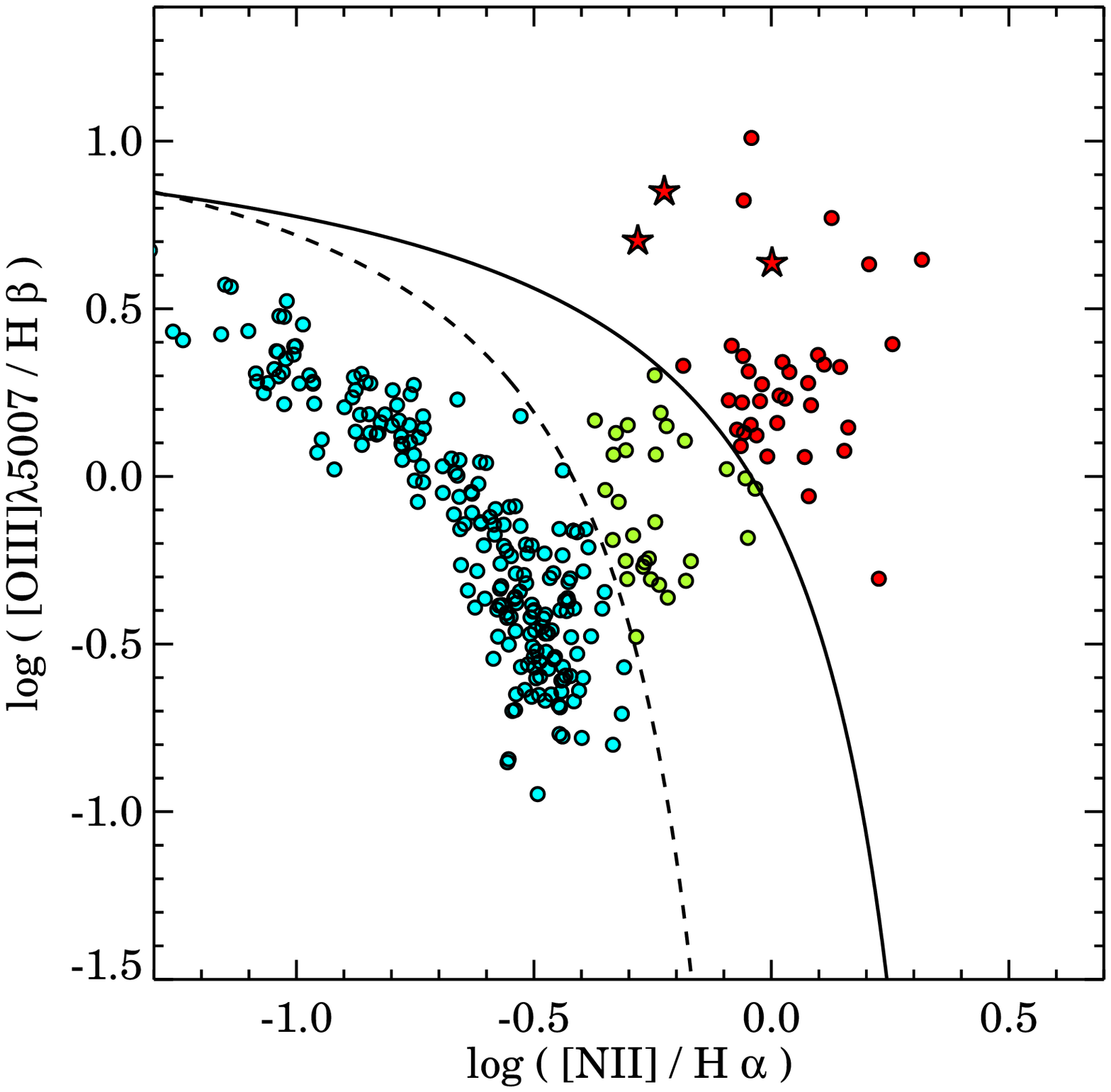}
\caption{BPT diagram for A2029 members. 
There are 277 emission-line objects:
 210 star-forming galaxies (cyan), 29 composite objects (green), and 38 AGNs (red). 
The stars indicate the {\it Chandra} X-ray point sources. }
\label{bpt}
\end{figure}

Figure \ref{bpt} shows the [OIII]/H$\beta$ versus [NII]/H$\alpha$ diagram for A2029 emission-line objects. 
The dashed and solid lines are the theoretical boundaries
 where purely star-forming \citep{Kauffmann03} and extreme star-forming galaxies \citep{Kewley01} 
 would appear, respectively. 
The objects with a high [OIII]/H$\beta$ and [NII]/H$\alpha$ ratio
 compared to the extreme star-forming galaxies are AGNs. 
We also classify star-forming and `composite' objects
 based on their relative positions with respect to the model lines.  
 
In A2029, there are 38 AGNs, 29 composite objects and 210 star-forming galaxies.  
The overall AGN fraction, i.e. $N_{AGN} / N_{member}$, is $\sim 3\%$. 
The AGN fraction is similar to
 the AGN fractions measured in individual clusters ($\sim 0.5 - 9\%$, \citealp{Deshev17,Habas18}). 
However, a direct comparison among the AGN fractions is not trivial 
 because the magnitude limit and spectroscopic completeness of other cluster redshift surveys vary. 
 
We measure the AGN fraction in fixed magnitude ranges 
 for a fairer comparison with cluster AGN fractions in the literature. 
\citet{Hwang12a} measure the AGN fractions from three volume-limited samples 
 extracted from stacked SDSS spectroscopic samples in eight clusters. 
The three volume-limited samples they used are:
 a bright sample with $-22.5 \leq M_{r} < -20.5$ and $0.04 \leq z \leq 0.1434$, 
 an intermediate-luminosity sample with $-20.5 \leq M_{r} < -19.5$ and $0.04 \leq z \leq 0.0927$, and
 a faint sample with $-19.5 \leq M_{r} < -18.5$ and $0.04 \leq z \leq 0.0593$. 
To compare with this result, 
 we estimate the AGN fraction in the bright magnitude range of $-22.5 \leq M_{r} < -20.5$. 
The AGN fraction, $\sim 10.0 \pm 1.7\%$, in the bright sample
 is similar to the bright sample of \citet{Hwang12a}, $\sim 7.6 \pm 0.3\%$. 
The AGN fractions of A2029 in the intermediate and faint magnitude ranges are negligible. 
 
Most A2029 AGNs (92\%) are located in galaxies with $M > 10^{10} M_{\odot}$; 
 14 ($37\%$) AGN host galaxies have $M > M_{\rm star} = 10^{10.7} M_{\odot}$ \citep{Sohn17a}. 
The mass distribution of A2029 AGN host galaxies differs from 
 that of A85 where a half of the AGNs reside in low-mass galaxies with $M < 10^{9.5} M_{\odot}$ \citep{Habas18}. 
\citet{Pimbblet13} examine the variation of AGN fraction as a function of host galaxy mass
 based on a stacked sample of six clusters. 
They show that more massive cluster galaxies are more likely to host AGNs.  
We find a similar trend; the AGN fraction is higher for A2029 members with larger stellar mass. 

To examine the radial dependence of the AGN fraction, 
 we compute the frequency of AGN among cluster members at various normalized clustercentric radii. 
There is no clear variation in the AGN fraction 
 within $R_{cl} < \sim 1.5 R_{200}$, 
 where the redshift survey is complete. 
The AGN fraction increases rapidly at $R_{cl} > 1.5 R_{200}$
 although the redshift survey is incomplete in this region. 
The radial dependence of the fraction of AGN and composite objects follow the same trend. 
The constant AGN fraction at the cluster center contrasts with 
 \citet{Pimbblet13} who found a significant decrease in the innermost region
 of six local clusters. 
\citet{Hwang12a} examine more details of the radial AGN fraction
 by computing the AGN fraction for early and late types, separately.  
The AGN fraction in early types shows a radial dependence; 
 the AGN fraction in late types changes little as a function of clustercentric distance. 
The majority of the AGN host galaxies in A2029 are late types ($\sim 66\%$);
 thus the weak radial dependence in AGN fraction seems consistent with the results from \citet{Hwang12a}. 
  
We also explore X-ray point sources in the A2029 field
 based on the {\it Chandra} source catalog (CSC) Release 2.0 \citep{Evans10}. 
The CSC provides a list of X-ray point sources identified based on
 {\it Chandra} ACIS and HRI observations. 
The CSC lists 26 X-ray sources within 40 arcmin of the A2029 center.  
There are 10 X-ray sources with spectroscopic counterparts within $3\arcsec$ of the X-ray center;
 there are four cluster members and six background sources with $z > 0.29$. 
Among the four X-ray emitting cluster members, 
 three are spectroscopic AGNs and one is an elliptical galaxy without emission-lines.
These three spectroscopic AGNs are luminous X-ray AGNs 
 with X-ray luminosities of $2.0 - 11.0 \times 10^{42}$ erg s$^{-1}$.  
Table \ref{chandra} lists the X-ray point sources in the A2029 field.

\begin{deluxetable*}{lccccccccc}
\tablecolumns{10}
\tabletypesize{\footnotesize}
\tablewidth{0pt}
\setlength{\tabcolsep}{0.05in}
\tablecaption{{\it Chandra} X-ray Point Sources in the A2029 Field}
\tablehead{
\colhead{{\it Chandra} ID}  & \colhead{} & \colhead{R.A.$_{X}$} & \colhead{Decl.$_{X}$} & \colhead{$f_{X}$\tablenotemark{a}} & 
\colhead{Object ID\tablenotemark{b}} & 
\colhead{$\theta_{\rm offset} (\arcsec)$\tablenotemark{c}} & 
\colhead{z} & \colhead{Membership} & \colhead{Spec. Type}}
\startdata
CXOJ151100.4+054921 & & 227.751875 &  5.82251 & $ 6.88^{+ 0.45}_{- 0.44}$ & 1237658780557836762 &  0.18 &   0.5463 & N & Galaxy \\
CXOJ151106.3+054122 & & 227.776623 &  5.68966 & $67.25^{+ 2.93}_{- 2.89}$ & 1237655744020087024 &  0.48 &   0.0807 & Y &    AGN \\
CXOJ151037.2+054814 & & 227.655344 &  5.80391 & $ 9.14^{+ 0.66}_{- 0.65}$ & 1237658780557771187 &  0.29 &   1.2232 & N & Quasar \\
CXOJ151123.4+054041 & & 227.847748 &  5.67819 & $ 5.63^{+ 0.73}_{- 0.73}$ & 1237655744020087171 &  1.21 &   0.9294 & N & Quasar \\
CXOJ151133.6+054546 & & 227.890377 &  5.76302 & $11.11^{+ 0.77}_{- 0.78}$ & 1237662268074033709 &  0.50 &   0.0847 & Y &    AGN \\
CXOJ151038.9+055329 & & 227.662125 &  5.89141 & $ 1.24^{+ 0.37}_{- 0.37}$ & 587736546849063315  &  2.54 &   0.2973 & N &   Comp \\
CXOJ151127.3+053943 & & 227.864084 &  5.66196 & $ 2.53^{+ 0.60}_{- 0.60}$ & 1237655744020087813 &  2.15 &   0.7881 & N &     NA \\
CXOJ151025.3+055026 & & 227.605677 &  5.84075 & $ 1.12^{+ 0.32}_{- 0.33}$ & 1237662268073902520 &  1.87 &   0.0745 & Y & Galaxy \\
CXOJ151045.9+055557 & & 227.691550 &  5.93273 & $ 0.05^{+ 0.64}_{- 0.05}$ & 587736546849064050  &  2.63 &   0.5669 & N & Quasar \\
CXOJ151141.2+051809 & & 227.921920 &  5.30258 & $34.67^{+ 4.96}_{- 4.92}$ & 1237662267537228077 &  0.09 &   0.0842 & Y &    AGN
\enddata
\tablenotetext{a}{X-ray flux in unit of $10^{-14}$ erg s$^{-1}$ cm$^{-2}$. }
\tablenotetext{b}{SDSS Object ID (DR12 or DR7) for the optical counterparts. }
\tablenotetext{c}{Offset between the X-ray source and the SDSS optical counterpart. }
\label{chandra}
\end{deluxetable*}

\subsection{The Mass-Metallicity (MZ) relation}\label{mzr}

We derive the mass-metallicity relation, hereafter the MZ relation, 
 for star-forming galaxies in A2029. 
The MZ relation of clusters is often measured 
 based on small sample of cluster galaxies or on stacked cluster samples 
 \citep{Ellison09, Petropoulou11, Robertson12, Gupta16}. 
\citet{Petropoulou12} investigate the MZ relation of four nearby clusters (Coma, A1367, A779, A634) 
 based on SDSS spectroscopy. 
Here, we derive the cluster MZ relation based on a large spectroscopic sample for a single cluster, A2029. 
 
Similar to the sample we use for AGN identification, 
 we select galaxies with line fluxes that have $S/N > 3$
 for H$\alpha$, H$\beta$, [NII]$\lambda6584$. 
We also require $S/N > 3$ in the [O II]$\lambda 3727$ flux measurement. 
\citet{Foster12} show that $S/N$ cuts on [O III]$\lambda5007$ may introduce a bias at high metallicity. 
Thus, we do not apply a $S/N$ cut on [OIII]$\lambda5007$. 
We identify 227 star-forming galaxies based on the BPT classification for this analysis. 
The number of star-forming galaxies increases slightly compared with the number (210) quoted in Section \ref{agn}
 because we do not exclude low [O III]$\lambda5007$ objects. 
 
We compute the metallicity of A2029 member galaxies 
 based on the \citet{Kobulnicky04} method. 
We first calculate $R_{23}$ and $O_{32}$ ratios: 
\begin{equation}
{\rm R}_{23} = \frac{{\rm [O II]}\lambda3727 + {\rm [O III]}\lambda4959 + {\rm [O III]}\lambda5007}{{\rm H} \beta}, 
\end{equation}
and 
\begin{equation}
{\rm O}_{32} = \frac{{\rm [O III]}\lambda4959 + {\rm [O III]}\lambda5007}{{\rm [O II]}\lambda3727}. 
\end{equation}
The MZ relation is often derived using equivalent widths (EWs) of the lines \citep{Kobulnicky04}. 
Here, we use line fluxes instead of EWs 
 to compare directly with the results from \citet{Wu17} 
 who examine the environmental dependence of the MZ relation based on line fluxes from SDSS spectra. 
We assume the flux ratio between [O III]$\lambda4959$ and [O III]$\lambda5007$ is 3 \citep{Osterbrock06}
 and simply use 1.33 times [O III]$\lambda5007$ when we sum the [O III] line fluxes. 
  
\citet{Kobulnicky04} determine metallicities, i.e. 12 + log (O/H), of galaxies
 based on the relative positions of the R$_{23}$ and O$_{32}$ indices 
 with respect to model grids.
Following their method, we derive the metallicities of A2029 star-forming galaxies. 
The intrinsic uncertainty of individual metallicity measurements is $\sim 0.1$ dex \citep{Kobulnicky04}. 

\begin{figure}
\centering
\includegraphics[scale=0.49]{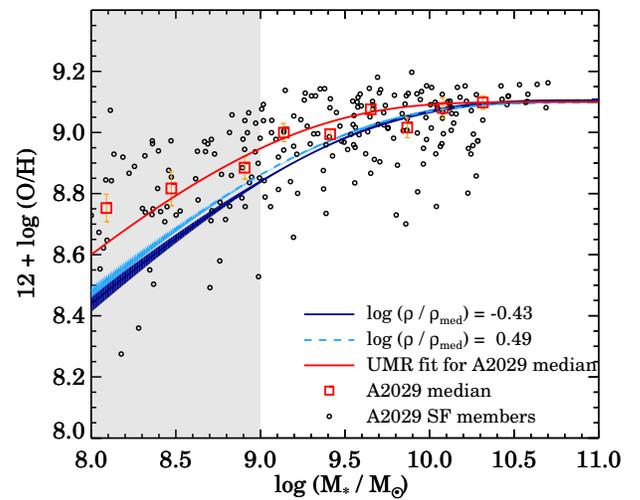}
\caption{12 + $\log (O/H)$ vs. stellar mass for A2029 members. 
Black circles show measurements for individual galaxies;  
 red squares show the median. 
The solid and dashed lines display the mass-metallicity relations 
 derived from SDSS galaxies in low-density region and high-density regions, respectively \citep{Wu17}. 
The shaded region indicates the mass range we exclude 
 when we compare the MZ relations from A2029 and with the SDSS galaxies. }
\label{mz}
\end{figure}

Figure \ref{mz} shows the MZ relation for A2029 star-forming galaxies. 
Black circles are individual measurements for the A2029 system members 
 and the red squares are the median metallicities as a function of stellar mass. 
We compare the A2029 MZ relation 
 with the MZ relation derived from SDSS galaxies at $z \sim 0.08$ in different density environments.  
\citet{Wu17} examine the MZ relation of SDSS galaxies 
 based on galaxies with $M > 10^{9} M_{\odot}$ in various density environments. 
The 8 Mpc kernel density they used, $\log (\rho/\rho_{med})$, is the sum of the weighted,
 kernel-smoothed luminosities of galaxies within 8 Mpc
 normalized by the median density of the SDSS main galaxy sample
 (see more details in \citealp{Wu17}).
In Figure \ref{mz}, 
 we show the SDSS MZ relations for $\log (\rho/\rho_{med}) \sim -0.48$ (light blue) and $\sim 0.43$ (dark blue) for simplicity.  
 
We fit the Universal Metallicity Relation (UMR) formulation of \citet{Zahid14a} 
 to the median metallicities of A2029 galaxies.  
The UMR formulation is :
\begin{equation}
12 + \log (O/H) = Z_{0} + \log \bigg[1 - exp \bigg(- \bigg[\frac{M_{*}}{M_{0}} \bigg]^{\gamma} \bigg) \bigg].
\end{equation}
In their study of environmental effects, \citet{Wu17} use this UMR formulation. 
Because the impact of local density on $Z_{0}$ and $\gamma$ is small, 
 they fit the MZ relations with fixed $Z_{0}$ and $\gamma$ to quantify the environmental effect on $M_{0}$.
Following their approach, we use a fixed $Z_{0} = 9.100$ and $\gamma = 0.505$. 
We note that the $Z_{0}$ and $\gamma$ change little if we do not fix them for fitting the UMR formulation. 
The red solid line shows the best-fitting model with $M_{0} = 8.86 \pm 0.04$. 
If we limit the fitting range to $M > 10^{9} M_{\odot}$ identical to \citet{Wu17}, 
 the $M_{0}$ is $9.05 \pm 0.05$. 
 
\citet{Wu17} show that 
 the best-fitting $M_{0}$ decreases significantly as the relative density increases. 
They also provide a relation between $M_{0}$ and the local density (eq (15) in \citealp{Wu17}). 
The 8 Mpc kernel density around A2029 is $\log (\rho/\rho_{med}) \sim 1$ (private communication with P.-F. Wu): 
 A2029 represents the highest density region in SDSS spectroscopic sample. 
The expected $M_{0}$ for the A2029 MZ relation based on the $M_{0}$-local density relation 
 is $9.12 \pm 0.03$, consistent with the $M_{0}$ we derive.    
 
The MZ relation for A2029 is similar to the SDSS MZ relation for the mass range $M > 10^{9.5} M_{\odot}$. 
We note that the MZ relation saturates in this mass range \citep{Zahid14a}.
In other words, the shape of the MZ relation in this mass range is insensitive to the sample redshifts and environments. 
 
Interestingly, 
 the A2029 population is more metal rich than the SDSS field population for $M < 10^{9.5} M_{\odot}$. 
This comparison is only valid to $M = 10^{9} M_{\odot}$ where the SDSS MZ relations were derived. 
This difference is more significant than 
 the difference among the SDSS field samples in different density environments.  

\citet{Petropoulou12} also find higher metallicity at low stellar mass
 for members of four nearby clusters including Coma.
In their sample, particularly Coma and A1367 galaxies, 
 the cluster galaxies with $10^{8} M_{\odot} < M_{*} < 10^{9} M_{\odot}$ within $R_{cl} < R_{200}$ 
 are metal-rich compared to their counterparts at $R_{cl} > R_{200}$. 
They suggest that the interaction between cluster galaxies and the intracluster medium 
 affects the metal content of low-mass galaxies in dense environments. 
Furthermore the intracluster medium tends to remove gas from these systems (e.g. \citealp{Gunn72}) 
 supporting the suggestion of \citet{Wu17} that these galaxies have less gas associated with them.
The A2029 MZ relation supports these suggestions based on the additional comparison with SDSS field samples. 

\subsection{E+A Galaxies}\label{ea}

An E+A galaxy has the spectrum of an elliptical galaxy 
 along with strong Balmer absorption lines typical of A type stars 
 \citep{Dressler83, Couch87}. 
These galaxies are probably a post-starburst population 
 where star formation was quenched within the last few Gyr
 \citep{Zabludoff96, Caldwell96, Tran03, Tran04, Goto05}.

The role of local environment in the formation and evolution of E+A galaxies remains unclear. 
Many studies examine the connection between local environment and the E+A phenomenon
 \citep{Dressler83, Zabludoff96, Dressler99, Blake04, Poggianti04, Quintero04, Goto05, Paccagnella17}.
At $z > 0.3$,
 E+A galaxies are more abundant in clusters than in the general field \citep{Tran03,Tran04}.
\citet{Lemaux17} show that 
 the post-starburst fraction is similar in field, group, and cluster environments at $z > 0.6$. 
However, they also show that 
 the ratio between the post-starburst and emission-line populations 
 is larger in clusters than in lower density regions. 
 
In the local universe ($z \sim 0.1$),
 E+A galaxies are predominantly in the field or in poor groups
 \citep{Zabludoff96, Blake04, Quintero04, Goto05}. 
A significant number of E+A galaxies were recently identified in local clusters:
 Coma \citep{Poggianti04} and A3921 \citep{Ferrari05}. 
\citet{Paccagnella17} provide a census of post-starburst galaxies
 based on the OmegaWINGS local cluster sample, 
 which includes 32 clusters at $0.04 < z < 0.07$.  
They suggest that the fraction of these post-starburst galaxies increases slightly
 as the clustercentric distance decreases. 
 
We identify E+A galaxies in A2029 
 with H$\delta$ EW $< -5$ \AA~ and [O II]$\lambda3727$ EW $< 2.5$ \AA~
 following the standard definition \citep{Goto03, Zahid16b}. 
A negative EW indicates absorption. 
The procedure we use to measure the H$\delta$ EW is outlined in \citet{Goto03}. 
We use the H$\delta$ (wide) definition: 
 the blue continuum is 4030-4082 \AA~ and 
 the red continuum is 4122-4170 \AA. 
The H$\delta$ EW is the sum of pixels in the line at 4082-4122 \AA. 
The details of H$\delta$ EW measure are described in \citet{Zahid16b}. 

\begin{figure}
\centering
\includegraphics[scale=0.49]{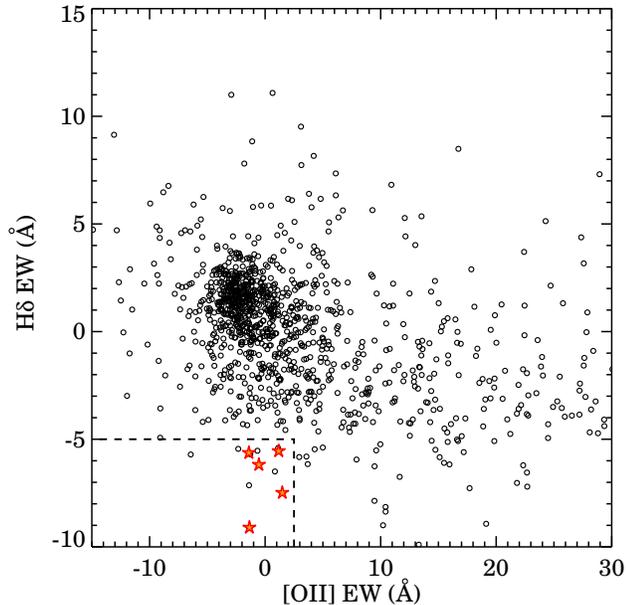}
\caption{
H$\delta$ EW vs. [O II]$\lambda3727$ EW for members of the A2029 system. 
Blue dashed lines indicate the boundary we use for identifying E+A galaxy candidates. 
Black circles are A2029 spectroscopic members and 
 red stars are the five E+A galaxies in A2029. }
\label{epa_ew}
\end{figure}

\begin{figure}
\centering
\includegraphics[scale=0.5]{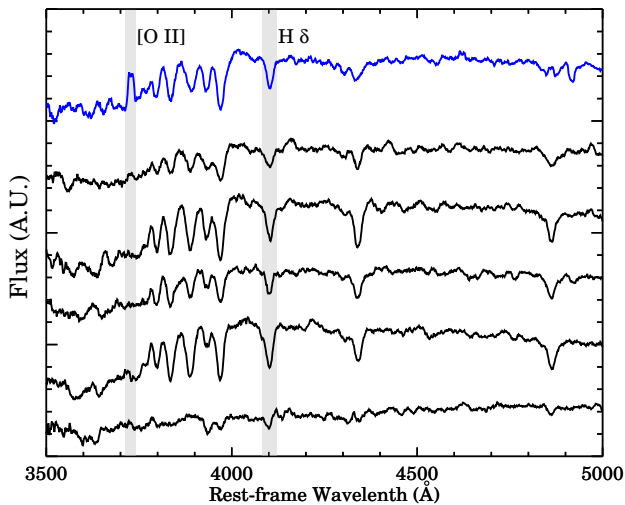}
\caption{ 
Rest-frame spectra of the five E+A galaxies (black) in the A2029 system. 
The fluxes of the spectra are arbitrarily shifted. 
The E+A galaxies are sorted by distance from the A2029 main cluster center:
 the top black spectrum is closest to the A2029 center. 
The spectrum at the bottom is from the E+A galaxy in SIG. 
The top spectrum shows a rejected E+A candidate with strong H$\delta$ absorption and emission lines. 
Note that this rejected E+A candidate has strong [OII] emission. }
\label{epa_spec}
\end{figure}

Figure \ref{epa_ew} shows 
 the H$\delta$ and [O II]$\lambda3727$ EW distribution for the A2029 system members. 
The dashed lines indicate the boundary we use for identifying 
 E+A galaxy candidates. 
There are 14 E+A galaxy candidates. 
We visually inspect the spectra of the E+A candidates
 to eliminate candidates with strong emission-lines. 
The galaxies with strong emission-lines are e(a) and A+em objects \citep{Poggianti99, Balogh99}, 
 which may have ongoing star-formation, or they may be AGN. 
We reject these E+A candidates with emission-lines from our E+A sample. 
We display the spectra of five E+A galaxies 
 along with an example spectrum of an E+A candidate with emission lines rejected during visual inspection (Figure \ref{epa_spec}). 

We identify four E+A galaxies within $R_{200}$ of the main cluster 
 and one E+A galaxy near SIG. 
The E+A galaxy in SIG shows weak E+A features and has a low stellar mass. 
Green diamonds in Figure \ref{spatial} mark the locations of these E+A galaxies. 
All of these E+A galaxies have projected radius $R_{cl} > 400$ kpc. 
The spatial distribution of the A2029 E+A galaxies suggests that 
 the E+A galaxies may not survive near the cluster core. 
We note that projection effects are important for assessing the true spatial distribution of the E+A galaxies. 
Although the E+A galaxies in A2029 are projected well within $R_{200}$, 
 their physical distances could be much larger than $R_{200}$.

\begin{figure}
\centering
\includegraphics[scale=0.40]{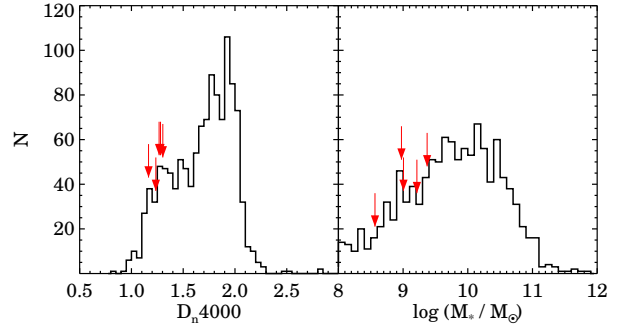}
\caption{
The $\dn$ (left) and stellar mass (right) distribution of A2029 system member galaxies. 
Red arrows show the five E+A galaxies in A2029. }
\label{epa_prop}
\end{figure}

Figure \ref{epa_prop} shows the $\dn$ and the stellar mass (M$_{*}$) distributions 
 of normal (histogram) and E+A (arrows) galaxies in A2029. 
E+A galaxies in different density environments and in different redshift ranges 
 generally have $\dn < 1.5$ \citep{Balogh99, Zahid16b}. 
The low $\dn$ reflects the post-starbust character of the E+A population.
The E+A galaxies in A2029 are less massive than 
 $M_{*} = 10^{10} M_{\odot}$ 
 (cf. M$_{*, A2029} \sim 5 \times 10^{10} M_{\odot}$, \citealp{Sohn17a}). 
The E+A galaxies in Coma and A3921 are also mostly faint ($M_{V} \geq -20$)
 and thus probably have low stellar mass. 
 
Historically, 
 ram pressure stripping has been considered a plausible mechanism 
 for turning off star formation in infalling galaxies.  
Thus, we compute the ram pressure exerted on E+A galaxies 
 based on the recipe from \citet{Gunn72}.  
When a galaxy falls into a cluster,
 it feels ram pressure due to the intracluster medium (ICM):
\begin{equation}
P_{ram} \approx \rho_{ICM} v^{2},
\end{equation}
where $\rho_{ICM}$ is the ICM density and 
 $v$ is the relative velocity of the galaxy with respect to the cluster mean. 
The restoring gravitational force per unit area 
 can be estimated \citep{Gunn72, Vollmer01}:
\begin{equation}
F = \Sigma_{gas} v_{rot}^{2} R_{disk}^{-1},
\end{equation}
 where $\Sigma_{gas}$ is gas surface density, $v_{rot}$ is rotational velocity of a disk, 
 and $R_{disk}$ is a disk radius. 
If the ram pressure exceeds the restoring gravitational force, 
 the gas in the disk may be removed.

To estimate the $\rho_{ICM}$, 
 we use the ICM density profile from \citet{Walker12}
 who measure the radial gas density within $R_{cl} < 20\arcmin$ 
 based on {\it Suzaku} X-ray observations. 
We fit a $\beta$ model with a form 
 $\rho_{g} = \rho_{0} [1 + (r / r_{c})^2]^{-\beta}$
 to their gas density (Figure 4 in \citealp{Walker12}). 
The best fit model parameters are 
 $\rho_{g} = 2.7 \times 10^{-26}$ g cm$^{-3}$, 
 $r_{c} = 0.88$ arcmin, and $\beta = 0.86$. 
The fitting parameters have large uncertainties, 
 but they are acceptable for our purpose. 
Based on the fit, we calculate the $\rho_{ICM}$ at the projected distance of A2029 E+A galaxies
 from the cluster center. 
We compute the ram pressure for each E+A galaxy using the $\rho_{ICM}$ 
 and the measured relative radial velocity with respect to the cluster mean.
 
\begin{figure}
\centering
\includegraphics[scale=0.49]{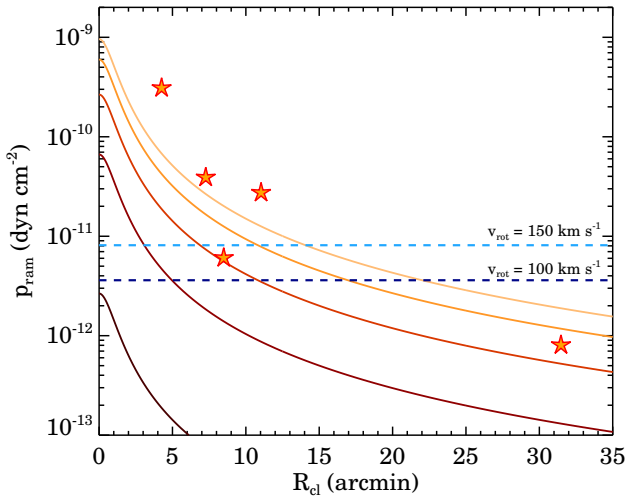}
\caption{
Ram pressure on a galaxy falling into A2029 
 with a relative velocity of 100, 500, 1000, 1500 and $2000~\kms$
 (solid lines from bottom to top)
 as a function of projected radial distance.
The horizontal dashed lines show the restoring gravitational force per unit area of a disk galaxy 
 with rotational velocity 100 (lower) and $150~\kms$ (upper).
Star symbols show the five E+A galaxies in A2029. }
\label{epa_ram}
\end{figure}

Figure \ref{epa_ram} shows the ram pressure on the five E+A galaxies 
 as a function of the projected distance. 
Solid lines in Figure \ref{epa_ram} plot
 the ram pressure from the ICM if a galaxy falls into A2029 
 with relative velocity 100, 500, 1000, 1500, and $2000~\kms$. 
Here, the ram pressure we compute is only indicative 
 because we use the projected clustercentric distance and the one-dimensional velocity. 
These effects may well compensate for one another. 

The dashed lines show 
 the restoring gravitational force per unit area for a typical disk galaxy 
 with rotational velocity 100 and $150~\kms$. 
We assume a gas surface density of $10^{-21}$ cm$^{-2}$ and disk radius of 5 kpc 
 \citep{Vollmer01, Lee17}.
 
Our simplified approach facilitates quantitative calculation of 
 ram pressure stripping and gravity on the E+A galaxies.  
\citet{Jaffe18} discuss caveats including
 projection effects in the orbit, 
 inclination angle, 
 and differences in disk models of the infalling galaxies. 
Resolving projection effects in the orbit is challenging. 
Thus, the ram pressure we calculate is only indicative of the possible role of ram pressure stripping. 

The comparison in Figure \ref{epa_ram} suggests that
 ram pressure is a possible explanation for all of the E+A galaxies in A2029.   
The ram pressure on the four E+A galaxies within the $R_{200}$ of A2029 
 exceeds the restoring gravitational force. 
For the E+A galaxy in SIG, 
 we probably underestimate the ram pressure 
 because we only compute the impact from the ICM around the main cluster
 using the relative motion of the galaxy with respect to the main cluster. 
Direct detection of the stripped gas would be a strong test of the ram pressure model (e.g. \citealp{Giovanelli85, Lee17, Jaffe18}).

\subsection{Central Velocity Dispersions}\label{vdisp}

The central velocity dispersion of a galaxy, reflecting stellar kinematics, 
 may be one of the key observables connecting the galaxy and its dark matter halo \citep{Schechter15, Zahid16}. 
Based on the Illustris simulation,
 \citet{Zahid18} show that the stellar velocity dispersion of the quiescent population is 
 tightly correlated with the dark matter halo velocity dispersion.
Thus, the stellar velocity dispersion is a good proxy for the dark matter halo properties.

\citet{Sohn17a} also discuss the power of the velocity dispersion to study cluster galaxy populations.
Unlike luminosity or stellar mass, 
 the velocity dispersion measurement is independent of systematic uncertainties in photometric data 
 (i.e. crowded field photometry or the dependence on stellar IMF or star-formation history of the stellar mass measurement).   
\citet{Sohn17a} measure the velocity dispersion function of quiescent members of the A2029 system.
The velocity dispersion function is a tool for studying the dark matter halo distribution
 in analogy to the luminosity and stellar mass functions. 
 
\begin{figure}
\centering
\includegraphics[scale=0.49]{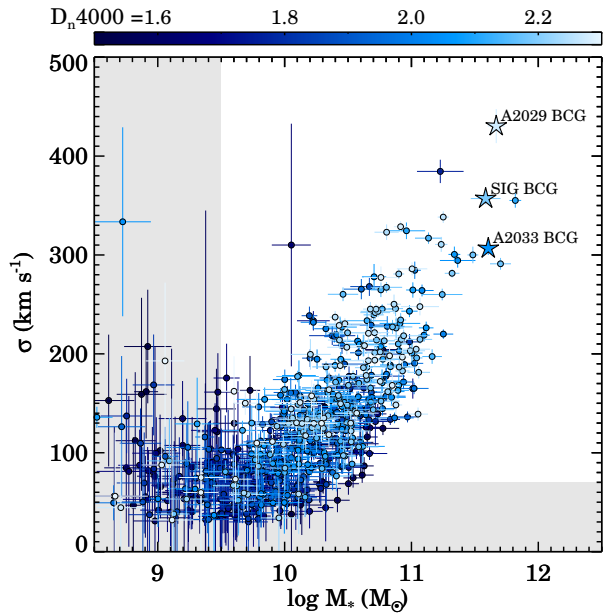}
\caption{
Velocity dispersion of A2029 member galaxies with $\dn > 1.5$ as a function of stellar mass. 
The colors of the symbols indicate $\dn$; light blue indicates high $\dn$ and dark blue indicates low $\dn$. 
The star symbols show the brightest galaxies of A2029, A2033, and SIG. 
The gray area indicates the region where either the velocity dispersion measurements are not robust or 
 where the spectroscopic survey is incomplete. }
\label{sigmass}
\end{figure}

Based on the updated catalog including new velocity dispersion measurements, 
 we examine the relation between velocity dispersion and stellar mass for the A2029 quiescent galaxies. 
Figure \ref{sigmass} shows the velocity dispersion of A2029 quiescent members as a function of stellar mass color coded by $\dn$. 
A2029 members with larger stellar mass generally have higher velocity dispersions.  
The higher $\dn$ galaxies tend to have higher velocity dispersions at a given stellar mass. 
These relations between velocity dispersion, stellar mass and $\dn$ are also observed 
 in Coma cluster members \citep{Sohn17a} and in the SDSS field galaxies \citep{Zahid16}. 

The star symbols in Figure \ref{sigmass} display
 the brightest galaxies of A2029 and the two infalling groups, A2033 and SIG. 
We use SDSS DR7 photometry of the BCGs when we compute their stellar masses.  
The brightest cluster galaxy (BCG) of A2029 (IC 1101) has the largest velocity dispersion, $\sigma = 430 \pm 17~\kms$ among the cluster members. 
The velocity dispersions of the brightest galaxies of SIG ($\sigma = 307 \pm 6~\kms$) and A2033 ($\sigma = 357 \pm 6~\kms$) are also very large. 

Interestingly, 
 the velocity dispersions of the brightest galaxies in A2029 and its subsystems 
 are roughly correlated with the masses of the systems. 
\citet{Sohn18b} estimate the velocity dispersions of the A2029, A2033 and SIG systems:
 $\sigma_{A2029} = 967 \pm 25~\kms$, $\sigma_{A2033} = 701 \pm 74~\kms$, and $\sigma_{SIG} = 745 \pm 62~\kms$. 
The relation between the central galaxies and the cluster velocity dispersions suggests 
 a connection between the brightest galaxy properties and the properties of the host system. 
 
\subsection{$\dn$ Distribution}\label{dn}

Despite its strength, 
 $\dn$ has not been widely used for studying galaxy populations. 
However, analyses based on $\dn$ add another powerful probe \citep{Balogh99, Luparello13}.
\citet{Balogh99} investigate the differential evolution of galaxies in 
 clusters and the field by tracing the $\dn$ distribution. 
Following this approach, 
 we explore the $\dn$ distribution of A2029 galaxies 
 to understand the evolution of galaxies within the cluster and the infalling groups. 

The impact of cluster environment on galaxy evolution is intensively studied 
 based on various tracers from multi-wavelength data. 
Many previous studies show evidence of environmental effects 
 including variation in color, morphological fraction, gas content, and (specific) star formation rates of galaxies 
 in varying density environments (e.g. \citealp{Oemler74, Rines05, Boselli06, Park07, Blanton09, Thomas10, Haines15, Barsanti18}). 
In general, galaxies in dense region are old and quiescent compared to their counterparts in less dense regions. 

\citet{Tyler13} examined environmental effects in A2029 
 based on the star formation activity of the cluster members. 
From {\it Spitzer} data and Hectospec spectra, 
 they identified 444 A2029 spectroscopic members with 24$\mu$m emission.
For these spectroscopic members, 
 they estimated star formation rates from the 24$\mu$m and Far UltraViolet luminosities. 
They also derived star formation rates based on H$\alpha$ equivalent widths from the Hectospec data. 
The star-forming galaxies in A2029 follow a star formation rate - stellar mass relation similar to field star-forming galaxies. 
The A2029 star-forming galaxies generally have higher star formation rates 
 compared to the star-forming galaxies in Coma with similar stellar masses. 
\citet{Tyler13} interpreted this difference
 as a marker of differing accretion histories of star-forming galaxies in A2029 and Coma.

The $\dn$ distribution of the entire A2029 system member galaxies is bimodal (Figure \ref{epa_prop} (a)). 
The population with $\dn \gtrsim 1.5$ consists of old (mean stellar age $> 1$ Gyr) and quiescent galaxies; 
 the other population consists of emission-line galaxies \citep{Woods10}.
The quiescent fraction is high ($f_{\dn > 1.5} = 69\%$) in A2029 
 as in other local clusters \citep{Balogh99, Deshev17}. 
The $\dn$ distribution of A2029 system members differs significantly from the $\dn$ distribution for field galaxies, 
 where the $\dn \leq 1.5$ population is dominant (e.g. \citealp{Mignoli05, Vergani08, Woods10}).
However, 
 comparison between the quiescent fraction in the field and clusters is not trivial, 
 because the $\dn$ index depends on a stellar mass \citep{Kauffmann03, Geller14, Geller16}. 
 
\begin{figure}
\centering
\includegraphics[scale=0.49]{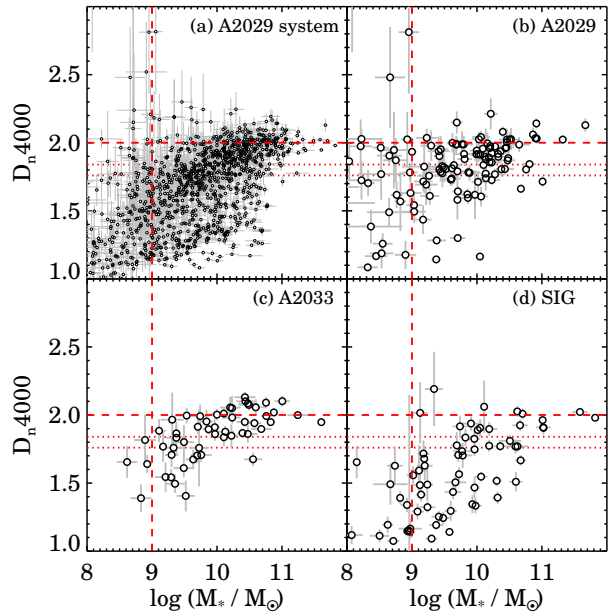}
\caption{ 
$\dn$ vs. stellar mass for members of (a) the A2029 system, (b) A2029 only, (c) A2033, and (d) SIG.
The panel (b)-(d) shows the members within $R_{proj} < 500$ kpc from the center of each system. 
The vertical dashed line shows $M = 10^{9} M_{\odot}$.
The dashed horizontal line is $\dn = 2.0$. 
The dotted horizontal lines show $\dn = 1.76$ and $\dn = 1.84$,
 where the potentially `accreted population' is located (Section \ref{trace} and Figure \ref{dn_sim}). }
\label{dn_mass}
\end{figure} 
 
Figure \ref{dn_mass} (a) plots 
 the $\dn$ of the members of the A2029 system as a function of stellar mass. 
The quiescent and star forming populations of A2029 members are clearly distinct in this plot. 
A2029 members show a $\dn$ dependence on stellar mass; 
 more massive galaxies tend to have larger $\dn$. 
A similar $\dn$ dependence on  stellar mass is observed 
 in other cluster (e.g. A520, \citealp{Deshev17}) and 
 field samples \citep{Geller14, Geller16, Haines17}. 
The fraction of quiescent galaxies in A2029 with $M < 10^{10} M_{\odot}$
 is as expected much larger than the fraction measured in the field at $0.1 < z < 0.2$ \citep{Geller16}. 
Although the field quiescent fraction is not measured at the exactly same redshift, 
 the higher quiescent fraction among less massive galaxies indicates that 
 galaxies in a dense environment are generally older or more rapidly quenched 
 compared to their counterparts in lower density regions.   

We plot the $\dn$-stellar mass relations for A2029, A2033, and SIG members in Figure \ref{dn_mass} (b) - (d). 
For fair comparison, we use member galaxies within $R_{proj} < 500$ kpc from the center of each system. 
We estimated that 
 there are 8 and 18 possible contaminating A2029 members at the distances of A2033 and SIG, respectively. 
To understand the impact of these A2029 contaminants, 
 we randomly sample 8 and 18 A2029 members in an annulus covering the A2033 and SIG regions. 
We excluded A2033 and SIG members when we sample the possible A2029 contaminants. 
We repeat this random sampling process 10000 times. 
The $\dn$ of the randomly sampled A2029 contaminants are uniformly distributed over the range of $1.0 < \dn < 2.2$. 
In other words, 
 the impact of contamination by A2029 members at the projected distances of A2033 and SIG has 
 a negligible effect on the $\dn$ distribution of the subsystem members.
 
The $\dn$ distributions of A2029, A2033, and SIG members differ significantly. 
The median $\dn$ of A2029 members with $10^{10} M_{\odot} < M_{*} < 10^{11} M_{\odot}$ is $\sim 1.92 \pm 0.16$
 and the less massive members have lower $\dn$. 
In the same mass range, the median $\dn$ of A2033 members is slightly larger, but within the uncertainty ($\sim 1.99 \pm 0.11$). 
More importantly, A2033 members show a very tight $\dn$ distribution around $\dn \sim 2.0$
 indicating that majority of the members were quenched at the same time. 
SIG members with similar mass show a much broader $\dn$ distribution. 
The differences in the $\dn$ distributions in the three subsystems suggest that 
 the galaxy populations in the A2029 subsystems have very different histories
 even though they are part of the same larger bound system.
 
\section{DISCUSSION}\label{discussion}
\subsection{A Picture of Galaxy Evolution in the A2029 System}

The dense spectroscopic survey of A2029 offers 
 a comprehensive view of the galaxy populations in A2029 
 based on a cleanly selected, large sample of members with little contamination by foreground/background galaxies. 
Taking advantage of the rich sample of A2029, 
 we can study the details of the galaxy properties in a single cluster.  
The $\dn$ distribution of A2029 system members (Section \ref{dn})
 provides a probe of the morphology-density relation in the cluster environment, 
 complementary to the usual considerations of star-forming galaxies. 
Because quiescent galaxies dominate clusters, 
 $\dn$ provides a denser tracer than the properties of the star forming populations. 
Here we discuss the implications of the $\dn$ distribution of A2029 members for the cluster evolutionary history. 

The $\dn$ index is a powerful age indicator of the stellar population. 
Aside from age, $\dn$ is affected by the metallicity \citep{Kauffmann03, Woods10}. 
For example, the $\dn$ of quiescent galaxies can vary up to $\sim 20\%$ for a large range of metallicity, 
 $0.8~Z_{\odot} - 2.5~Z_{\odot}$ \citep{Kauffmann03}. 
However, the metallicity of quiescent galaxies in a single cluster often has only a small variation \citep{Rakos07}. 
Thus, metallicity probably has little impact on the interpretation of the $\dn$ distribution of  A2029 system members. 

\begin{figure}
\centering
\includegraphics[scale=0.49]{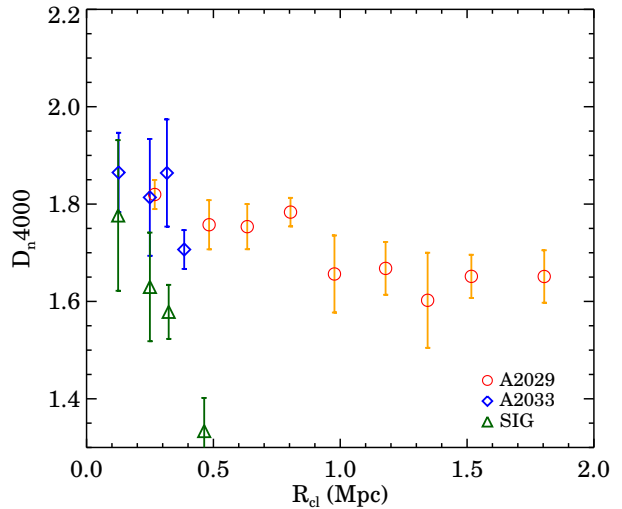}
\caption{ 
Median $\dn$ as a function of (normalized) clustercentric distance
 from the center of A2029 (red circles), A2033 (blue diamonds) and SIG (green triangles). }
\label{dn_grad}
\end{figure}

Figure \ref{dn_grad} displays the median $\dn$ of the members of A2029, A2033, and SIG
 as a function of projected distance from the cluster (group) centers.
We estimate the median $\dn$ using A2029 members within $R_{200}~(= 1.91$ Mpc) and 
 the A2033 and SIG members within 500 kpc, respectively.
The dashed circles in Figure \ref{spatial} indicate the boundaries we use for investigating the $\dn$ variation.  
We only use the members with $\log (M_{*} / M_{\odot}) > 9.0$ when we calculate the median $\dn$. 
 
The median $\dn$ of the galaxies in the primary cluster A2029 (red circles) declines as a function of projected distance. 
There is a slight fluctuation at larger radius perhaps resulting from the contamination by members of infalling groups.
The decline of $\dn$ is consistent with the decreasing mean $\dn$ in the outskirts of clusters 
 derived from a stacked sample of 15 clusters at $z \sim 0.3$ \citep{Balogh99}. 

Remarkably, 
 the median $\dn$s of the galaxies in the two infalling groups A2033 (blue diamonds) and SIG (green triangles)
 also decrease as the projected distances from their respective centers increases.
The decline of median $\dn$ for A2033 members follows the trend for A2029 members. 
SIG members show a similar decline, 
 but the median $\dn$s are much lower than for A2029 and A2033 members. 

Figure \ref{dn_grad} has two important implications.  
The $\dn$ gradients in the A2029 systems are consistent with 
 the general picture of environmental effects (quenching) on galaxy evolution 
 (e.g. \citealp{Oemler74, Dressler80, Balogh99, Christlein05, Rines05, Boselli06, Peng10, Rasmussen12, Haines15, Barsanti18}). 
The impact of local environment on galaxy evolution depends on galaxy mass \citep{Peng10}.
Because we limit our sample to $\log (M/M_{\odot}) > 9.0$, 
 where the survey is complete, the $\dn$ gradients in the A2029 systems are robust.
In addition, 
 we derive the median $\dn$ variation among the three A2029 subsystems 
 using the galaxies in two stellar mass bins  $9.0 < \log (M_{*} / M_{\odot}) \leq 10.0$ and $10.0 < \log (M_{*} / M_{\odot}) \leq 11.0$. 
The $\dn$ gradients are the same 
 but with larger uncertainties due to the smaller number of objects.   

The decreasing $\dn$ in the two infalling groups as a function of projected distance 
 suggests that galaxies are `processed' within the group environment and prior to infall into the main cluster.  
Previous studies investigate the fraction of star forming galaxies or star formation activity
 in galaxy groups (e.g. \citealp{Balogh04, Wilman05, Jeltema07, McGee11, Rasmussen12}). 
They show the decline of star formation activity (or the fraction of star forming galaxies) in the group centers.
$\dn$ provides a denser tracer of this galaxy processing based on the entire group population
 including both star-forming and quiescent populations.
 
Projection effects and contamination by interlopers with high radial velocities 
 are important in interpreting the analyses of cluster populations \citep{Rines05}. 
Combining their extensive redshift surveys with a projected Navarro-Frank-White profile, 
 \citet{Rines05} estimate the fraction of ``infall interlopers," 
 galaxies within $R_{proj} < R_{200}$, but $R_{3D} > R_{200}$. 
They suggest that at least 20\% of non-emission line galaxies and 50\% of emission-line galaxies are infall interlopers. 
The A2029 sample includes these infall interlopers, certainly. 
Indeed, some emission-line galaxies with large relative radial velocities lie near the center of A2029. 
These galaxies most probably lie within the extended infall region 
 because the A2029 system is so well separated from the foreground/background in redshift space. 

The median $\dn$ distribution is a robust probe of the environmental effect in spite of projection effects. 
The \citet{Rines05} estimate indicates contamination by ``infall interlopers" is more significant for the star-forming population
 (or lower $\dn$ population). 
If we could remove contamination by the interlopers, 
 a larger fraction of star-forming galaxies would be excluded from the sample than quiescent galaxies. 
Thus, the median $\dn$ gradient is likely to be even steeper. 
  
Identifying the physical mechanisms for environmental effects is not straightforward. 
Gravitational interactions among galaxies may disturb the gas dynamics in galaxies resulting in 
 either star formation or rapid consumption of gas. 
Hydrodynamic interaction between galaxies and baryonic matter in their environment 
 can also be responsible for differential evolution of galaxies in dense environments
 (e.g. ram pressure stripping \citealp{Gunn72}). 
Presumably, all of these physical processes work together 
 and result in the differential evolution of galaxies in a cluster.
Regardless of the physical mechanism, 
 the results we derive based on complete spectroscopy suggest that 
 these evolutionary processes are important throughout the A2029 system. 
  
\subsection{Tracing the Merger History of A2029}\label{trace}

The X-ray sloshing pattern in A2029 indicates that the dynamical history of A2029 is complex. 
This sloshing pattern in a galaxy cluster can form as a result of a merger with a subcluster \citep{ZuHone10}. 
In the deep {\it Chandra} observations, 
 the X-ray sloshing pattern of A2029 is obvious \citep{Clarke04, PaternoMahler13}. 
Comparison between the observed A2029 sloshing pattern and hydrodynamic simulations \citep{ZuHone10} 
 suggests that A2029 accreted a subcluster about 2-3 Gyr ago \citep{PaternoMahler13}. 
We compare this timescale with other indicators of the cluster history. 
We refer all considerations to the age of the universe at $z_{A2029} = 0.078$. 
 
We check for evidence of a feature in the $\dn$ index corresponding to the accretion time
 suggested by the X-ray sloshing pattern. 
We assume that member galaxies of the accreted subcluster 
 became quiescent galaxies ($\dn = 1.5$) when the subcluster accreted by A2029 produced the sloshing pattern.
We then compute the `current' $\dn$ of these galaxies 
 based on the assumption that the subcluster interaction occurred either 2 or 3 Gyr ago. 

We simulate the time evolution of $\dn$ for a galaxy after star formation is quenched 
 (see \citealp{Zahid15}). 
We assume a model quiescent galaxy with solar metallicity 
 which forms stars with a constant rate for 1 Gyr before quenching. 
For this model galaxy, 
 we construct a synthetic spectrum of a quiescent galaxy 
 using the Flexible Stellar Population Synthesis (FSPS; \citealp{Conroy09, Conroy10}). 
$\dn$ is measured from the spectrum as the model galaxy ages. 

\begin{figure*}
\centering
\includegraphics[scale=0.49]{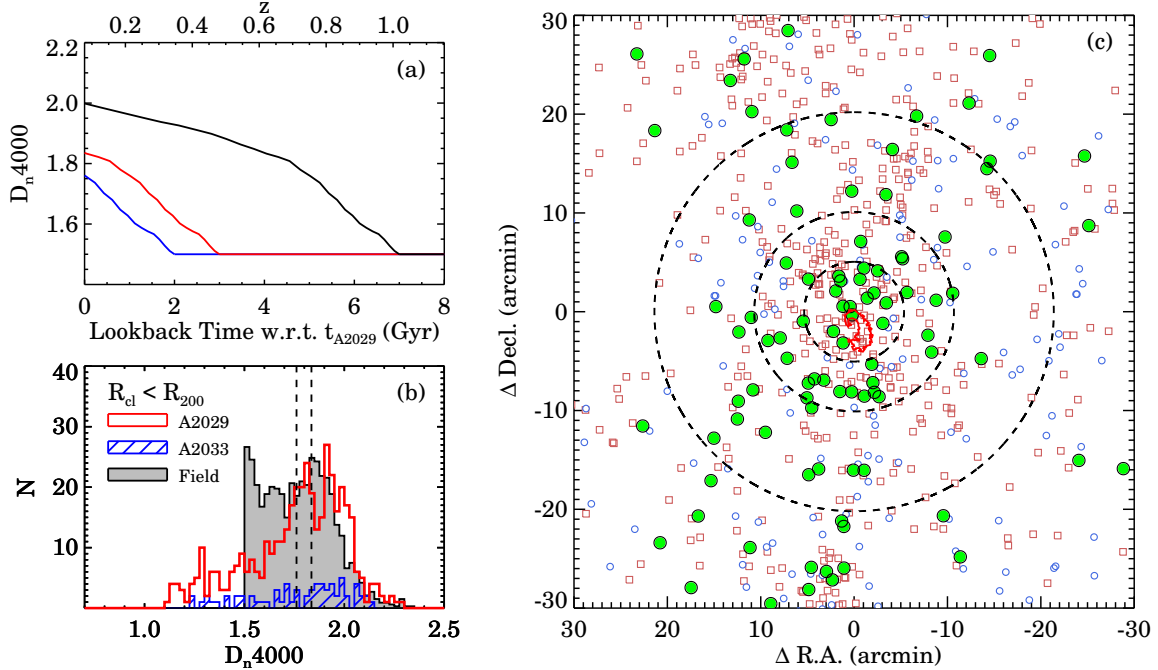}
\caption{ 
(a) Time evolution of $\dn$. 
Blue, red and black lines show the $\dn$ evolution of a model galaxy 
 if it became quiescent 2, 3, and 7 Gyr ago with respect to the age of the universe at $z_{A2029}$, respectively. 
(b) The $\dn$ distribution of A2029 members with $\log (M/M_{\odot}) > 9.0$ within $R_{cl} < 1.91$ Mpc $ = R_{200}$ (open histogram). 
The hatched histogram is the $\dn$ distribution of A2033 members within the same mass range within $R_{cl} < 500 kpc$. 
The filled histogram is the field $\dn$ distribution
 obtained from the SHELS \citep{Geller14} and the hCOSMOS \citep{Damjanov18a}. 
Field galaxies have $\dn > 1.5$, $M_{*} > 10^{9} M_{\odot}$, and $0.0 < z \leq 0.2$.
The vertical dashed lines mark the $\dn$ of the model galaxies which became quiescent 2 and 3 Gyr ago. 
The comparison suggests that the excess of A2029 members at $1.76 < \dn \leq 1.84$ is significant. 
(c) Spatial distribution of the A2029 members with $1.76 < \dn \leq 1.84$ (green star symbols). 
Red squares and blue circles are A2029 members with $\dn > 1.5$ and $\dn \leq 1.5$, respectively. 
The red line displays the X-ray sloshing pattern from {\it Chandra} observations \citep{PaternoMahler13}. 
The dashed circles indicate $R_{cl} = 0.25, 0.5$ and $1.0 R_{200}$. }
\label{dn_sim}
\end{figure*}

Figure \ref{dn_sim} (a) shows the time evolution of $\dn$. 
Red and blue lines show the change of $\dn$ as function of time for a galaxy that became quiescent 2 and 3 Gyr ago, respectively. 
For comparison, we show the epoch when a galaxy with $\dn = 2.0$ became quiescent (the black line, $\sim 7$ Gyr ago). 
The galaxies that became quiescent 2 and 3 Gyr ago have $\dn = 1.76$ and $\dn = 1.84$, respectively. 
In other words, in a simple accretion picture, 
 the galaxies that accreted onto A2029 when the interaction occurred should have $\dn$ in between 1.76 and 1.83. 

Figure \ref{dn_sim} (b) displays the $\dn$ distribution of A2029 members
 with $\log (M/M_{\odot}) > 9.0$ within $R_{cl} < R_{200}$ (= 1.91 Mpc). 
We limit the mass range to minimize systematic effects due the incompleteness of the survey. 
Interestingly, there is an excess at $1.76 < \dn < 1.84$ 
 (indicated by the vertical lines, as well as the dotted horizontal lines in Figure \ref{dn_mass}) 
 separate from the dominant peak at $\dn \sim 2.0$. 
The $\dn$ distribution of A2033 members (the hatched histogram) does not show a similar excess. 
The $\dn$ excess of the A2029 member distribution is consistent with the idea that
 there was an accretion event 2 to 3 Gyr ago. 
 
To test the significance of the $1.76 < \dn < 1.84$ peak further, 
 we compare the cluster $\dn$ distribution with the field $\dn$ distribution.  
We obtain the $\dn$ sample from the Smithsonian Hectospec Lensing Survey (SHELS) F1 and F2 survey \citep{Geller14}. 
We also compile the $\dn$ distribution from the hCOSMOS survey \citep{Damjanov18a}, 
 a dense MMT/Hectospec spectroscopic survey of $r \leq 21.3$ galaxies in the COSMOS field. 
These dense surveys at $z < 0.2$ are complete to the similar mass limit ($M \gtrsim 10^{9} M_{\odot}$) comparable with the A2029 survey. 
Thus, we limit the field sample within $0.0 < z \leq 0.2$. 
The $\dn$ distributions of the field population are essentially identical for $ z \lesssim 0.6$. 
 (see Figure 8 in \citealp{Geller14}). 
The field comparison sample includes 4259 galaxies; 1442 have $\dn > 1.5$.
 
The filled histogram in Figure \ref{dn_sim} (b) displays 
 the $\dn$ distribution of the field galaxies. 
To reduce selection effects, 
 we compute the $\dn$ distribution based on randomly sampled field galaxies 
 and take the median from the 10000 realizations. 
We plot the $\dn$ distribution with $\dn > 1.5$ for simplicity; 
 the young population with $\dn \leq 1.5$ dominates the field $\dn$ distribution. 
We also normalized the field $\dn$ histogram 
 using the peak of the A2029 $\dn$ distribution in order to compare the shapes of the $\dn$ distributions. 

The field $\dn$ distribution differs significantly from the cluster $\dn$ distribution. 
The field $\dn$ distribution is fairly flat. 
There is a slight excess at $1.8 < \dn < 1.9$ near the secondary peak of the A2029 $\dn$ distribution. 
However, the field $\dn$ distribution does not show a dip at $\dn \sim 1.9$
 underscoring the significance of the secondary peak of A2029 members. 
 
Figure \ref{dn_sim} (c) plots the spatial distributions of 
 the `accreted population' (green star symbols) within the narrow $\dn$ range $1.76 < \dn < 1.84$.
Red squares and blue circles are other members with $\dn > 1.5$ and $\dn \leq 1.5$. 
The red line shows the X-ray sloshing pattern taken from Figure 7 in \citet{PaternoMahler13}.
The `accreted population' is located outside the sloshing pattern 
 and shows a strong concentration toward the cluster center. 

The consistency between the time scales estimated based on the $\dn$ distribution and the X-ray sloshing pattern 
 is a fascinating clue to the cluster history.
The $\dn$ excess at the central region of A2029 suggests that 
 the $\dn$ distribution indeed traces the subcluster interaction that produced the sloshing pattern. 
The combination of detailed X-ray data and dense spectroscopy provides 
 a potentially rich platform for the exploration of cluster merger histories.  

\subsection{Three Subsystems, Three Different galaxy populations}

The most striking feature in our spectroscopic survey of the A2029 infalling groups 
 is their very different galaxy populations (Figure \ref{dn_mass}). 
Quiescent galaxies dominate in A2033 and star-forming galaxies dominate in SIG. 
They also differ substantially from the main cluster,
 which contains members with a broader $\dn$ distribution than for A2033 members and larger median $\dn$ than SIG members. 
 
The difference in the $\dn$ distributions of A2033 and SIG 
 does not result from the different subsystem masses
 because the total masses of A2033 and SIG do not differ significantly. 
Furthermore, the stellar mass distributions of the member galaxies are similar (Figure \ref{dn_mass}).
The local galaxy number densities in A2033 and SIG are also comparable; 
 in fact, the number density of SIG is slightly higher than for A2033 in the central region. 
This comparison suggests that there is a large stochastic component in the evolution of members of galaxy subsystems of similar mass.  
 
The X-ray properties of A2033 and SIG provide an additional demonstration
 of the marked differences between them. 
Although A2033 and SIG have similar masses measured from multi-probes including velocity dispersion and weak lensing, 
 the X-ray properties are very different. 
A2033 is bright in the X-ray with a high temperature of 3.7 keV; 
 SIG is not very bright in X-ray and its morphology is disturbed. 
The disturbed morphology of SIG indicates that SIG may not be dynamically relaxed, 
 possibly explaining the dominant young population among SIG members. 

\citet{Sohn18b} examine the evolutionary history of the A2029 system 
 using the two-body model \citep{Beers82} to trace the accretion history of A2033 and SIG.
They show that A2033 and SIG are gravitationally bound within the A2029 system. 
Based on the two-body model solution, SIG will accrete onto A2029 within $\sim 2.3$ Gyr. 
A2033 seems to be moving away from the primary cluster A2029, 
 but it may accrete within $\sim 2.8$ Gyr given the uncertainties in cluster mass and radial velocity. 

The impact of accretion of A2033 and SIG on the galaxy population of A2029 will differ significantly
 because of their totally different populations. 
As Figure \ref{dn_mass} (b) and (c) show, 
 galaxies in A2029 and A2033 are old and quiescent.  
When A2033 falls onto A2029, only quiescent galaxies will be added to the original A2029 population. 
Star formation activity is already suppressed. 
This merger would be undetectable with the argument we apply to explore the accretion event that produced the sloshing pattern.
In contrast to the accretion of A2033, 
 the accretion of SIG will supply a younger population (star forming galaxies) to A2029. 
The difference between the median $\dn$s of SIG and A2029 members within $R_{cl} < 500$ kpc is $\sim 0.18$ 
 roughly corresponding to an effective age difference of $\sim 2$ Gyr. 
This merger could be detectable with the argument we apply to the sloshing pattern origin.

Our result suggests that 
 stochastic effects are important in the evolution of galaxy populations in massive clusters.
The impact of the accretion of an infalling group on the cluster galaxy population 
 obviously depends on the galaxy population in the infalling systems.  
Sometimes, galaxies in the infalling group are already processed 
 as much as or more than the cluster galaxies. 
In other accreting systems, there is a younger population 
 which may include a substantial fraction of star-forming galaxies.  
Combining star formation activity with the dense tracer 
 $\dn$ enable a more nuanced understanding of the composition of groups and clusters 
 including age differences among the quiescent populations. 

\section{CONCLUSION}\label{conclusion}

Based on a dense and complete spectroscopic survey, 
 we examine the physical properties of galaxies in the local massive cluster A2029.
The A2029 system is unusually rich with 1215 spectroscopic members. 
This rich sample enables a study of the cluster population within a single cluster
 thus avoiding the common stacking technique. 
We examine the census of A2029 system members. 

The AGN fraction in A2029 is 3\%, consistent with the previous studies of AGN fraction in similarly massive systems.  
Most A2029 AGNs are bright ($M_{r} \leq -20.5$), massive ($M_{*} > 10^{10} M_{\odot}$) galaxies. 
Three of A2029 AGNs have X-ray counterparts in the {\it Chandra} X-ray point source catalog. 

We derive the stellar mass-metallicity (MZ) relation of A2029 and
 compare it with the SDSS field MZ relations \citep{Wu17}.  
For $M_{*} > 10^{9.5} M_{\odot}$, 
 the A2029 MZ relation is essentially identical to the field MZ relations. 
Interestingly, 
 A2029 star-forming galaxies tend to have higher metallicity than 
 the field galaxies in the mass range $10^{9.0} M_{\odot} < M_{*} \leq 10^{9.5} M_{\odot}$.  
This excess metallicity is also observed in Coma cluster.  
Interaction between these lower mass galaxies and the intracluster medium
 may affect their metal content. 
 
We identify five E+A galaxies in A2029;
 four of them are within $\sim$ 400 kpc of primary cluster center;  one is near the southern infalling group (SIG). 
These E+A galaxies have low $\dn~(< 1.5)$ and low stellar mass ($10^{8.5} M_{\odot} < M_{*} < 10^{9.5} M_{\odot}$).  
The ram pressure for these objects probably exceeds the restoring gravitational force \citep{Gunn72}.

\citet{Sohn18b} investigate the structure of A2029 
 based on this dense spectroscopy combined with X-ray and weak lensing data. 
They identify at least two subsystems, A2033 and SIG, in the infall region of A2029. 
These subsystems are gravitationally bound to A2029 and they will possibly accrete onto A2029 in a few Gyr. 

To explore connections between the galaxy populations in the main body of A2029 and the two infalling groups, 
 we use the $\dn$ index, an age indicator. 
Remarkably, the $\dn$-stellar mass relations of A2029, A2033, and SIG differ significantly. 
In both A2029 and A2033, 
 galaxies with $M_{*} > 10^{10} M_{\odot}$ have higher $\dn \sim 2.0$. 
The $\dn$ distribution of the A2029 members is broader than for A2033 members at a given stellar mass. 
The tight $\dn$-stellar mass relation of A2033 members suggests that A2033 members became quiescent galaxies at the same time. 
In contrast SIG, 
 a subsystem with a mass similar to A2033, has the broadest $\dn$ distribution and
 contains members with generally lower $\dn$ than either A2029 and A2033. 
SIG members are the youngest population in the A2029 system. 
Thus, the future accretion of the aged A2033 and the young SIG promise a totally different impact on the resulting population of the main cluster.  

The X-ray sloshing pattern of A2029 indicates that 
 A2029 probably experienced a subcluster merger a few Gyr ago \citep{PaternoMahler13}. 
If galaxies became quiescent ($\dn \sim 1.5$) around the time of the merger,
 the $\dn$ distribution could show an excess coincident with the merger epoch.
Remarkably, 
 the $\dn$ distribution of A2029 members does show an excess at the expected $\dn \sim 1.8$ .
The consistency of time scales traced by the independent $\dn$ distribution and X-ray sloshing pattern suggests
 that the $\dn$ distribution derived from dense spectroscopic samples of individual massive clusters could be a useful tracer of past mergers. 
 

\acknowledgments
We thank Perry Berlind and Michael Calkins for operating Hectospec and 
 Susan Tokarz for helping with the data reduction. 
We thank Po-Feng Wu for kindly providing the kernel density measurements of A2029 galaxies and for helpful discussion.
We also thank Antonaldo Diaferio, Ian Dell'Antonio, and Ken Rines for fruitful discussions.
This paper uses data products produced by the OIR Telescope Data Center, 
 supported by the Smithsonian Astrophysical Observatory.  
J.S. gratefully acknowledges the support of the CfA Fellowship. 
The Smithsonian Institution supported the research of M.J.G. and 
 H.J.Z. is supported by the Clay Postdoctoral Fellowship. 
This research has made use of NASA's Astrophysics Data System Bibliographic Services. 
Observations reported here were obtained at the MMT Observatory, a joint facility of the University of Arizona and the Smithsonian Institution.
Funding for SDSS-III has been provided by the Alfred P. Sloan Foundation, 
 the Participating Institutions, the National Science Foundation, 
 and the U.S. Department of Energy Office of Science. 
The SDSS-III web site is http://www.sdss3.org/. 
SDSS-III is managed by the Astrophysical Research Consortium for 
 the Participating Institutions of the SDSS-III Collaboration including 
 the University of Arizona, the Brazilian Participation Group, 
 Brookhaven National Laboratory, University of Cambridge, 
 Carnegie Mellon University, University of Florida, the French Participation Group, 
 the German Participation Group, Harvard University, the Instituto de Astrofisica de Canarias, 
 the Michigan State/Notre Dame/ JINA Participation Group, Johns Hopkins University, 
 Lawrence Berkeley National Laboratory, Max Planck Institute for Astrophysics, 
 Max Planck Institute for Extraterrestrial Physics, New Mexico State University, 
 New York University, Ohio State University, Pennsylvania State University, 
 University of Portsmouth, Princeton University, the Spanish Participation Group, 
 University of Tokyo, University of Utah, Vanderbilt University, University of Virginia, 
 University of Washington, and Yale University.

\bibliographystyle{apj}
\bibliography{ms}

\clearpage

\end{document}